\documentstyle[prb,aps,epsfig,floats,aps,multicol,fleqn,eqsecnum]{revtex}

\newcommand{\lessim} {\mathop{\,<\kern - 1.05 em \lower 1.ex \hbox {$\sim$}\,}}
\newcommand{\grtsim} {\mathop {\,> \kern - 1.05 em \lower 1.ex \hbox 
{$\sim$}\,}}
\begin{document}
\normalsize \draft \title{Weakly 
correlated electrons on a square lattice: a renormalization group theory}

\author{D. Zanchi}
\address{Institut f\"ur Theoretische Physik der Freien
Universit\"at Berlin, Arnimallee 14, 14195 Berlin, Germany\\
and \\
Laboratoire de Physique Th\'eorique et Hautes Energies. 
Universit\'es Paris VI Pierre et Marie Currie -- Paris VII Denis Diderot, 
2 Place Jussieu, 75252 Paris C\'edex 05, France. 
\footnote{present address}}

\author{H. J. Schulz\footnote{deceased}}
\address{Laboratoire de Physique des Solides, Universit\'e Paris-Sud,
91405 Orsay, France}
%\date{\today }
\maketitle

\widetext

\begin{abstract}
We formulate the exact Wilsonian renormalization group for a system of 
interacting fermions on a lattice. The flow equations for all vertices of 
the Wilson effective action are expressed in form of the Polchinski equation.
We apply this method to the Hubbard model on a square lattice using both 
 zero-- and  finite--temperature methods. 
Truncating the effective action at the sixth term in fermionic variables
we obtain the one--loop functional renormalization equations for the effective
interaction. We find the  temperature of the instability $T_c^{RG}$ as 
function of doping. 
We calculate furthermore the renormalization of the angle--resolved
correlation functions for the superconductivity (SC) and for the 
antiferromagnetism (AF).
The dominant component of the SC correlations is of the type 
$d_{x^2-y^2}$ while the AF fluctuations are of the type $s$.
Following the strength of both SC and AF fluctuation along the instability 
line we  obtain the phase diagram. The temperature
$T_c^{RG}$ can be identified with the crossover temperature $T_{co}$ 
found in the underdoped regime of the high--temperature superconductors, 
while in the overdoped regime $T_c^{RG}$ corresponds to the 
superconducting critical temperature.

\end{abstract}


\bigskip
PAR/LPTHE/98-60 \hspace{100mm} December 1998
\bigskip

%\begin{multicols}{2}

%\narrowtext

\newpage

\section{Introduction}

In systems of correlated fermions on a lattice some interesting and 
also puzzling physics seems to happen when interaction--induced localization
tendencies, antiferromagnetic, and superconducting fluctuations get mixed.
The standard example of such a system are the copper--oxide
superconductors. \cite{revHTC} In the underdoped regime, between the AF and
SC phases, correlations of both AF and SC type are strongly enhanced, and a
pseudogap is visible in the one particle spectrum and in the spin response
functions.  The pseudogap regime is limited from above by a crossover
temperature $T_{co}(x)$, a monotonously decreasing function of doping.  At
the temperatures $T\grtsim T_{co}$ the underdoped materials are ``strange
metals'': many physical properties are unlike those of a standard Fermi
liquid\cite{FL}.  In the overdoped regime $T_{co}(x)$ merges with the
critical temperature for superconductivity and the regime $T > T_{c}$ is
merely a Fermi liquid.  Another interesting feature of the phase diagram is
the unusual form of the order parameter.  After a rather long period of
controversies the $d_{x^2-y^2}$--symmetry is finally generally
accepted.\cite{d-wave} This is one of the reasons to believe that the
pairing mechanisms are tightly related to the antiferromagnetic tendencies
and not to the standard phonon--exchange mechanisms.  The
$d_{x^2-y^2}$--form of the superconducting correlations subsists also in the
pseudogap regime as is seen in recent angle--resolved photoemission
\cite{ARPES_psgap} and tunneling \cite{tunel_psgap} experiments.  The
simultaneous existence of strong AF correlations, as seen by NMR
\cite{NMR} or neutron--scattering \cite{neutrons} experiments, and even
localization tendencies as flattening of the band\cite{flat_band} make us
conclude that the interpretation of this regime only terms of 
superconducting or antiferromagnetic fluctuations only is not sufficient,
especially because we expect that they are coupled.

It is striking that some other apparently completely different systems of
correlated fermions have very similar properties.  A phase diagram with the
superconducting phase in the vicinity of the spin density wave
(i.e. antiferromagnetic) instability characterizes also the
quasi--one--dimensional Bechgaard salts \cite{JerS} and the
quasi--two--dimensional organic superconductors of the ET
family,\cite{Hadrienetal} where instead of doping the relevant parameter for
the phase diagram is pressure. However the common feature of all these
compounds is that they are systems of correlated fermions with reduced
dimensionality $(D<3)$ and with strongly anisotropic and more or less nested
Fermi surfaces.  The main points for the understanding of the three groups
of compounds are: (i) the destruction of the nesting by doping (cuprates) or
by applying pressure (Bechgaard salts and ET); (ii) the suppression of the
Umklapp processes by doping the the half-filled band (in the cuprates and
ET-s) or by making the half--filled band effectively quarter--filled through
the breaking the longitudinal dimerization by pressure (in the Bechgaard
salts).  Concerning the Bechgaard salts it is interesting to remark that
some very recent interpretations of the phase diagram of the $(\rm
TMTSF)PF_6$ material \cite{Gabay98} suggest that the intermediate regime
between the high--temperature 1D behavior and the low temperature 3D physics
is a strange 2D--liquid with properties very similar to those of the
underdoped cuprates above the crossover temperature $T_{co}$.

From the theoretical point of view it is certainly interesting to construct
a theory able to treat on the same footing antiferromagnetic and
superconducting 
tendencies in more than one dimension and
to follow how the result changes with some external parameter that destroys
nesting and the Mott--like localization.  The first question one can ask is
whether a purely repulsive model like for example the Hubbard model (or some
generalization of it) contains the coexisting and inter--depending
antiferromagnetic and superconducting correlations.  In such a model the
antiferromagnetic fluctuations are associated with the enhancement of the
particle--hole (p-h) propagators at low energies and the superconducting
tendencies appear through particle--particle (p-p) propagators. The Hubbard
model is appropriate because at half filling already a simple mean--field
calculation gives an antiferromagnetic instability at a finite
temperature. However if one tries to include also the p-p processes already
in the weak coupling limit the problem becomes nontrivial already in the
weak coupling limit: a simple mean--field theory is not able to follow both
p-h and p-p correlation channels. One can of course try to remedy this
problem by including summation of selected subseries of higher--order
diagrams. 
One such attempt was to calculate the effective Cooper
amplitude as a sum of  bubble and ladder 
 RPA series\cite{Berk_66,Scalapino_95}. 
The resulting Cooper amplitude is then used as  coupling constant for
assumed effective BCS theory. This procedure thus explicitely decouples
three diferent summations (RPA bubbles, RPA ladder, and BCS ladder)
without real justification.
The FLEX (conserving)
calculations \cite{bickers_89}  based on the similar simplifications
are also prejudiced by the choice of diagrams to be summed. 
The only way to
proceed systematically is to construct the renormalization group that takes
into account all p-p and p-h loops of a given order (or to use the
equivalent parquet approach). In {(quasi--)} one dimension the renormalization
group has been successfully used and is one of the basic theoretical
ingredients in the physics of low--dimensional metals
\cite{JerS,Rev1,Bourbonnais}.

In two dimensions only a limited number of simplified cases was solved by
the renormalization group.  The poor man's scaling applied only to the
interactions between electrons placed at the van Hove points it gives an
antiferromagnetic instability at half-filling and superconductivity of
$d_{x^2-y^2}$ symmetry if the deviation of the chemical potential $\mu$ from
its value at half-filling becomes of the order of critical temperature of
the antiferromagnetic state.\cite{HJS} The equivalent parquet approach has
been used for half-filling (but without the limitation to the van Hove
points) and also finds an antiferromagnetic instablity.\cite{dzy,DzYak} 
Parquet calculations for simple flat Fermi surfaces \cite{Zheleznyak}
give an antiferromagnetic instability but can not provide a continuous phase
diagram as function of some imperfect nesting parameter or band filling: the
$d_{x^2-y^2}$--like superconducting pole appears simply by cutting the p-h
part of the flow, as in ref.\onlinecite{HJS}. The scaling approach to a system
with a Fermi surface with both flat and curved parts \cite{doucot_97} has
also reported a superconducting instability in a purely repulsive model,
together with deviations from Fermi liquid behavior. However, the complete
one--loop renormalization group (or parquet) for the real {\it band} of
electrons with imperfect and tunable nesting, or doping, still remains
unresolved.  The main difficulty is related to the correct treatment of the
coupling between p-p and p-h channels.

Different authors have tried to avoid to take into account the coupling
between the different renormalization channels making drastic
simplifications or limiting themselves to some particular forms of the Fermi
surfaces or only to the low energy effective action.  In our former
publications \cite{ZS_prb,ZS_zfp} we have shown that in the Hubbard model
one can treat the p-h channel perturbatively if the filling is sufficiently
far from one half.  Then the renormalization group gives only a weak
Kohn--Luttinger like pairing. The p-p part of the flow is decoupled from the
p-h one in the low energy regime for the simple reason that the p-h part is
negligible there. Other calculations based on the perturbative treatment of
the p-h channel were also reported.\cite{gonzales_97,Ruwalds} If the Fermi
surface is well (but imperfectly) nested, and this is exactly the
interesting regime, this strategy does not work any more because both p-p
and p-h loops are non--perturbatively large, even for weak interactions. In
the case of the square Fermi surface (with or without the van Hove
singularities) taking only the leading logarithmic part the coupling between
the channels into account\cite{Kwon} is equally insufficient.  Another way
to proceed is to see the 2D Hubbard system as an ensemble of coupled
chains:\cite{Balents97} this approach gives a phase diagram with
superconductivity formed by pairs of electrons on different chains, giving
rise to a spatially anisotropic version of an $d$-wave order parameter.
Among the number of the theoretical approaches to the Hubbard model other
then via the loop--summations the Monte Carlo calculations take in principle
``everything'' into account but it is still unclear whether they give
\cite{Fettes}
 or not \cite{Scalapino_95,Zhang} the
superconductivity.

In the present paper we search to reliably determine the phase diagram of
the Hubbard model in the vicinity of half--filling where p-h processes are
non--perturbatively enhanced and at least nearly as important as p-p ones.
We also detect the dominant components of the angle--resolved correlation
functions for antiferromagnetism and superconductivity as function of
temperature. This allows us to know the symmetry of the microscopic fields
whose fluctuations become important. The method that we will use is a
generalization of Shankar's renormalization--group approach \cite{Shankar}
to an arbitrary form of the Fermi surface. In particular, the
Kadanoff--Wilson mode elimination (developed by Shankar for 2D fermions)
applied to the effective action
 with only two--particles interaction keeps only the
strictly logarithmic contributions to the flow. 
Thus even if the nesting is
very good but not perfect the p-h part of the flow would be zero because the
logarithmic singularity is destroyed by imperfect nesting. To keep the p-h
part of the flow finite even in the case of imperfect nesting we start by
formulating the {\it exact} Kadanoff--Wilson--Polchinski renormalization
group for fermions on a lattice.  It was formulated previously
\cite{Polchinski84,Morris94} only for the quantum fields with {\it one}
zero--energy point in the momentum space, like the $\phi^4$ field theory
(critical phenomena). In many--fermions system in more than one dimension we
have on the contrary the whole Fermi surface that plays the role of the zero
energy manifold, what makes the calculations more complicated.  Starting
with the full bandwidth as the initial energy cutoff we perform iterative
mode elimination reducing the cutoff $\Lambda$ around the Fermi surface.
Collecting at each step of the renormalization all the terms (cumulants) of
the effective action we obtain the Polchinski equation for the vertices of
the effective theory at the given step of the renormalization.  It is
important that even if the initial interaction was only a four--point
function (two--particle interaction) vertices of {\it all} higher orders are
created by the renormalization procedure.  Once the exact renormalization
group is formulated we proceed with its truncation at the one--loop level:
the one--loop truncation of the flow for the four--point vertex is done by
neglecting all renormalization--group--created vertices of order larger than
six. Shankar\cite{Shankar} already has remarked that the six--point function
created by the mode elimination is essential to get non--logarithmic
contribution to the four point vertex (the effective interaction). The
one-loop renormalization of the interaction that we obtain in this way
appears to be generally {\it non--local} in $\Lambda$, i.e. the flow of the
vertex at a given step of the renormalization depends on the values of the
vertex at former steps. This is certainly not a pleasant but (as far as we
can see) necessary property of the KWP procedure if we want to keep more
than just the purely logarithmic contributions.

We than apply the one--loop KWP renormalization group to the Hubbard model.
One further approximation we make is to consider the effective interaction
as a function only of the projection of the momenta to the square Fermi
surface (marginal interactions) while the radial dependence and dynamics are
neglected because they are irrelevant with respect to the Fermi--liquid
scaling.\cite{Shankar} We neglect also the renormalization of the
self--energy. If we take only marginal interactions into account this is
justified at the one--loop level because the renormalization of the weight
and of the lifetime of the electrons receives a nonzero contribution only at
the two--loop level. We thus renormalize only the interaction
$U(\theta_1,\theta_2,\theta_3)$, a function of three angular variables
corresponding to the angular parts of the three external momenta, the fourth
being determined by momentum conservation. We allow the $\theta$--variables
to be anywhere on the almost square Fermi surface and not only in the
configurations that give perfect nesting or zero center--of--the--mass
momentum: these two classes of the configurations would correspond only to
the processes with the leading logarithmic renormalization in the p-h and
p-p channels, respectively.  As will become clear later, taking all three
$\theta$--variables without constraints is the essential point of the
calculation because the coupling between p-p and p-h channels is appears
mostly through interactions that have other than just leading--logarithmic
flow.  This is a special feature of the square or almost square Fermi
surface and can be handled only by the non--local (outer--shell)
contributions to the flow, using the Polchinski equation.

The first aim of our calculation is to find the temperature at which the
system flows towards strong coupling. We associate this temperature with a
mean--field like critical temperature and call it $T_c^{RG}$.  A typical
mean--field theory then is regularized for $T<T_c$ by adding counterterms
that contain fermions bilinearly coupled to some order parameter. In our
theory the order parameter is not known a priory: it is determined by the
manner in which the {\it function} $U(\theta_1,\theta_2,\theta_3)$ diverges
at $T=T_c^{RG}$. We perform a detailed analysis of the behavior of the
angle--resolved correlation functions for antiferromagnetism and
superconductivity and obtain the type and the symmetry of the order
parameter determining the dominant correlations near the $T_c^{RG}$.  The
final result is the analogue of a mean--field phase diagram of the Hubbard
model. 
We
are considering a two--dimensional system where one should be careful about
the interpretation of $T_c$: in the case of magnetism, this indicates the
onset of well--defined finite--range correlations. For weak interactions,
this is typically a very well--defined crossover.\cite{schulz_89} In the
case of pairing $T_c^{RG}$ can be identified with the onset of 
quasi--long--range
order. However, in  real systems like copper oxides
even a weak inter--plane two--particle 
hopping (particle--hole--pair hopping for antiferromagnetism or Josephson 
tunneling for superconductivity)  stabilizes  a 3D long--range order.

In section II we begin by the formulation of the many fermion system on a
lattice in terms of functional integrals.  We introduce the concept of the
effective action and show how it can be formally calculated using the
partial trace technique. We than derive the Kadanoff--Wilson--Polchinski
exact renormalization group as one possible strategy for calculating the
effective action in terms of the renormalization group flow of all vertices.
Truncating the effective action at the level of sixth order vertices we
obtain the one loop renormalization group equations for the effective
interaction and for the selfenergy.  In section III we apply the
zero--temperature one--loop renormalization group to the Hubbard model on a
square lattice.  We derive the flow equations for the effective interaction
function and for the angle--resolved correlation functions of
superconducting and antiferromagnetic type.  After discretization of the
angle $\theta$ on the Fermi surface we integrate numerically the flow and
present the resulting phase diagram. In section IV we introduce finite
temperature explicitly in the renormalization group equations.  We then
calculate the fixed point values of the correlation functions at
temperatures near the instability.  The conclusions are given in section V.

\section{Formulation of the renormalization group for a many--fermion 
problem on a lattice}

The simplest model for 
interacting fermions on a two--dimensional square lattice is the 
Hubbard Hamiltonian 
\begin{equation} \label{Hubbarddir}
H=-t\sum _{\langle i,j\rangle ,\sigma} (a^{\dagger}_{i,\sigma }a_{j,\sigma
}+a^{\dagger}_{j,\sigma }a_{i,\sigma})+\frac{U_0}{2}\sum _in_in_i-\mu\sum
_in_i
\end{equation}
where $a_{i,\sigma }(a^{\dagger}_{i,\sigma })$ is the creation 
(annihilation) operator of an electron at the site $i$ with spin $\sigma$,
 $t$ is the
inter-site transfer integral, $\mu$ is the chemical potential, and $U_0$ is
the on-site Coulomb repulsion. After Fourier transform, the Hamiltonian writes
\begin{equation} \label{Hubbard}
H=\sum _{\sigma {\bf k}} \xi _{\bf k}
{a}^{\dagger}_{\sigma {\bf k}}
{a}_{\sigma {\bf k}}+
\frac{1}{2}\sum _{\sigma }
\sum _{{\bf k}_1,{\bf k}_2,{\bf k}_3}U_0
{a}^{\dagger}_{-\sigma,{\bf k_1+k_2-k_3}}
{a}_{-\sigma  {\bf k_2}}
{a}^{\dagger}_{\sigma {\bf k_3}}
{a}_{\sigma {\bf k_1}}\; ,
\end{equation}
where 
\begin{equation} \label{dispersion}
\xi _{\bf k}=-2t(\cos k_x + \cos k_y)-\mu
\end{equation}
and the momenta are
within the first Brillouin zone.
In this section we want to derive the renormalization group for a more general
problem. For that purpose we allow $\xi _{\bf k}$ to have a
 general dependence on ${\bf k}$. 
Furthermore, we suppose that the interaction can be nonlocal and 
dynamical, 
that is, we suppose that it depends on energies and momenta of the 
interacting particles.

The statistical mechanics of souch general  model is given by the 
partition function
\cite{Shankar}
\begin{equation} \label{partfunct}
Z=\int {\cal D}\bar{\Psi}{\cal D}{\Psi} \; e^{S\{ \Psi\} } 
\end{equation}
where the functional integration is over Grassmann variables 
$\bar{\Psi}(\Psi)$
for all electrons in the Brillouin zone.
The action S is given by
$$
S\{ \Psi \} =S_0\{ \Psi \}+S_I\{ \Psi \}=
T\sum_{\omega_{n}}
\sum _{\sigma
{\bf k}}  \bar{\Psi}_{\sigma K} (i\omega
_n   
-\xi _{\bf k}) {\Psi}_{\sigma K}+
$$
\begin{equation} \label{act}
+\frac{1}{2}\sum _{\sigma \sigma '}
T^3\sum_{\omega_{n_1},\omega_{n_2},\omega_{n_3}}
\sum 
_{\bf k_1,k_2,k_3}
U_0(K_1,K_2,K_3)
\bar{\Psi}_{\sigma K_3}\bar{\Psi}_{\sigma 'K_4}{\Psi}_{\sigma 'K_2}
{\Psi}_{\sigma K_1} 
\; .
\end{equation}
The variables $\bar{\Psi}(\Psi)$ are labeled by the energy--momentum vector
$K=(\omega _n,{\bf k})$.  $\xi _{\bf k}$ is the bare spectrum measured
from the Fermi level:
$$\xi _{\bf k}=\epsilon _{\bf k}-\mu\; ,$$ where $\epsilon _{\bf k}$ is
the band dispersion and $\mu$ the chemical potential.  The energies and
momenta are conserved so that $K_4(K_1,K_2,K_3)=(\omega _{n_1}+\omega
_{n_2}-\omega _{n_3}, {\bf k_1+ k_2 - k_3})$.  $U_0(K_1,K_2,K_3)$ is the
most general spin independent interaction, a function of the frequencies and
momenta. The derivation of the action (\ref{act}) for a general model is 
equivalent to the derivation for the Hubbard model \cite{Negele,Fradkin}, 
provided that
we  put $U_0(K_1,K_2,K_3)$  instead of the constant $U_0$ and keep 
$\xi _{\bf k}$ general.

We want to derive the low energy effective action (LEEA) for this model. The
low energy modes are the electronic degrees of freedom close to the Fermi
surface. We will use this criterion and use the energy variable $\xi _{\bf
k}^0$ to discriminate fast (high--energy) modes $\Psi _>$ from the slow
(low--energy) ones $\Psi _<$.  Let's choose some arbitrary nonzero high
energy cutoff $\Lambda$ defining a shell of wavevectors around the Fermi
surface.  The electronic variables can then be written
\begin{equation}\label{slow-fast}
\Psi _{\sigma, K}=\theta (|\xi _{\bf k}|-\Lambda)\Psi _{>,\sigma, K}+
\theta (\Lambda -|\xi _{\bf k}|)\Psi _{<,\sigma, K}.
\end{equation}
The slow modes are inside the shell $\pm \Lambda $ while the fast ones are
outside, with $|\xi _{\bf k}|$ going up to the physical cutoff $\Lambda
_0$ taken to be equal to the bandwidth so that we are sure that the whole
Brillouin zone is taken into account.  Note that the cutoff is imposed only
on momentum space, while the Matsubara frequencies remain unlimited. The
LEEA $S_{\Lambda}\{ \Psi _<\} $ is an action containing only slow modes and
gives the same partition function as $S$ eq.(\ref{act}), or formally
\begin{equation} \label{eff.act}
Z=\int  {\cal D}\bar{\Psi}_<{\cal D}{\Psi _<} \; e^{S_{\Lambda} \{ \Psi _<
\} } ,
\end{equation}
This means that $S_{\Lambda}\{ \Psi _<\} $ is calculated by taking the {\it
partial trace} over only fast modes in eq.(\ref{partfunct}):
\begin{equation}\label{LEEA}
S_{\Lambda}\{\Psi _<\}=\ln
\int  {\cal D}\bar{\Psi}_>{\cal D}{\Psi _>} \; e^{S \{ \Psi _< , \Psi _>
\} }
\end{equation}
The LEEA contains a new effective kinetic part $S_{0\Lambda}$ with a finite
selfenergy term and a new interaction $S_{I\Lambda}$.  We have {\it chosen}
that the Fermi surface for the bare electrons plays the role of the zero
energy manifold. This still does not mean that we make the approximation of
a Fermi surface unrenormalized by interactions: even if the Fermi surface of
the LEEA is different from the bare one we are still allowed to use $\xi _{\bf
k}^0$ for the bookkeeping of our mode elimination.

If we consider the slow modes as parameters, expression (\ref{LEEA}) can be
evaluated, at least formally, using the linked cluster theorem.\cite{Negele}
The result is composed of three terms:
\begin{equation} \label{LEEA1}
S_{\Lambda}\{\Psi _<\}=S\{\Psi _<\} +\Omega _> + \delta S\{\Psi _<\}
\end{equation}
Only the interaction part $S_I$ of the action $S$ can mix slow and 
fast modes:
\begin{equation}\label{action_slow_fast}
S_I=S_I\{\Psi _<\} +S_I\{\Psi _>\} +S_I\{\Psi _<,\Psi _>\}
\end{equation}
while $S_0$ is diagonal and can contain only one kind of modes 
in the same term:
\begin{equation}\label{action1_slow_fast}
S_0=S_0\{\Psi _<\} +S_0\{\Psi _>\}\; .
\end{equation}

The first term in the equation (\ref{LEEA1}) is then only a constant from
the point of view of the fast electrons and is equal to $S_0\{\Psi _<\} +
S_I\{\Psi _<\}$.  $\Omega _>$ is the grand potential (times $\beta$) of the
fast electrons as if they were decoupled from the slow ones:
\begin{equation}\label{omega}
\Omega _> = -\sum _{\bf k>} \ln (1+e^{-\beta \xi _{\bf k}})+\sum 
(\mbox{all connected clusters with}\; S_I\{\Psi _>\} )
\; .
\end{equation}
This term gives only a shift of the total free energy of the system. 

\begin{figure}
\centerline{\epsfig{file=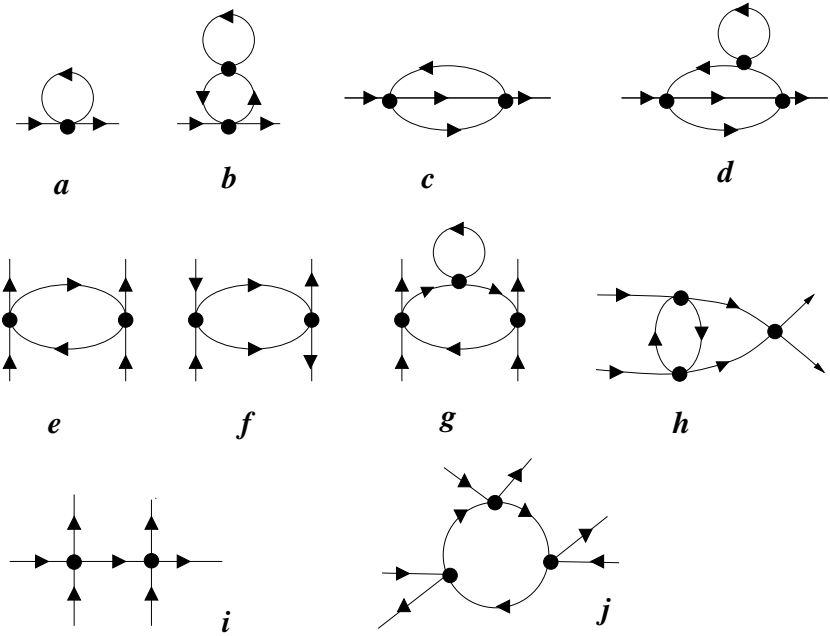,width=11cm}}
\caption{A few lowest order cumulants for 
$\delta S\{\Psi _<\}$. All internal lines are integrated only over the 
fast $(>)$ modes.}
\label{Fig1}
\end{figure}

The term $\delta S\{\Psi _<\}$ in eq.(\ref{LEEA1}) is the most interesting
one.  It brings the corrections due to the scattering processes of the slow
modes on the fast ones into the LEEA and is given by the sum of all
connected graphs composed of $S_I\{\Psi _>\} +S_I\{\Psi _<,\Psi _>\} $.  If
we draw the slow modes as external legs, the diagrams for $\delta S\{\Psi
_<\}$ are the clusters with at least two legs.  A few low order diagrams
for $\delta S\{\Psi _<\}$ are given in fig.\ref{Fig1}.
The terms with
two external legs, labeled  by $a$, $b$, $c$, 
and $d$ in figure \ref{Fig1}  
are the selfenergy terms, renormalizing $S_0$. The terms
with four legs: $e$, $f$, $g$ and $h$ 
 renormalize the quartic interaction term $S_I$.  The  terms
with six ($i$ and $j$) and more legs are new! They are {\it created} 
by the mode
elimination procedure.
 
Of physical interest is the LEEA for the electrons in the very vicinity of
the Fermi surface ($\Lambda \ll E_F$).  Even if the coupling is small some
of the loop diagrams will attain at low temperature $(T<\Lambda)$ large
values depending of the form of the Fermi surface.  For example, if the
Fermi surface is not close to van Hove singularities the particle--particle
(p-p) diagram ($f$ in fig. \ref{Fig1}) 
for the four--point vertex with zero center of the mass
momentum always has a logarithmic dependence like $\log (\Lambda_0/\Lambda
)$, where $\Lambda_0$ is the initial cutoff equal to the bandwidth.
If the Fermi surface is nested the
particle--hole (p-h) diagram ($e$ in fig. \ref{Fig1}) 
at $2{\bf k_F}$ behaves in the same way.  In
the Hubbard model close to half--filling the van Hove singularities make
both loops squares of logarithms. The perturbative calculation of the
expansion (\ref{omega}) for small $U_0$ is thus not straightforward: at
least some of sets of diagrams, containing {\em both p--p and p--h
subdiagrams}, have to be summed entirely. 
The lowest order diagram of that kind is the one denoted by
 $h$ in fig. \ref{Fig1}.
On the other hand the truncation
of the LEEA at fourth order is in general allowed for weak coupling.
However the direct summation of cumulants for $\delta S\{\Psi _<\}$ (like
T-matrix or RPA summation) can be performed in a useful and controlled way
only for a limited number of physical problems, that is when some subsets of
diagrams are dominant. The direct parquet summation for a general Fermi
surface in more than one dimensions is probably very hard.  It has been done
only for the case of perfectly nested (flat) Fermi surface
\cite{DzYak,Zheleznyak}.

The problem is even more difficult if the coupling is not small.  Then the
criteria of most important sets of diagrams are not clear any more and even
the truncation of LEEA at quartic or sextic term in $\Psi _<$ is not
justified any more.

\subsection{Kadanoff--Wilson--Polchinski 
renormalization group: Exact formulation} 
A tractable way to construct the
{\it exact} LEEA is to use the Kadanoff--Wilson--Polchinski renormalization
group.  Let us call the initial cutoff (the bandwidth) $\Lambda _0$ and
parameterize $\Lambda $ by the renormalization parameter $l$ so that
$\Lambda = \Lambda _0 \exp(-l)$.  The idea of the renormalization group is to
consider the transformation $S\equiv S_{\Lambda _0} \rightarrow S'_{\Lambda
_0 \exp (-l)}$ as an infinite set of infinitesimal mode eliminations
\begin{equation} \label{successive}
S_{\Lambda _0} \rightarrow S^{(1)}_{\Lambda _0 \exp (-dl)} \rightarrow
S^{(2)}_{\Lambda _0 \exp (-2dl)} \rightarrow ... 
\rightarrow S'_{\Lambda _0 \exp (-l)}
\end{equation}
At each step we eliminate $ \Lambda dl$  of modes at a
distance $\Lambda$ from both sides of the Fermi surface.  We will see that
the mode elimination of an infinitesimal shell of degrees of freedom is much
simpler than the one--step procedure discussed in the previous section.

From now on we will call the LEEA simply the effective action because in the
process of successive mode elimination (\ref{successive}) $\Lambda $ can
have any value between $\Lambda _0$ and zero.  Indeed, it is of physical
interest to follow the flow of the effective action $S_ {\Lambda}$ as
${\Lambda}$ decreases.

\begin{figure}
\centerline{\epsfig{file=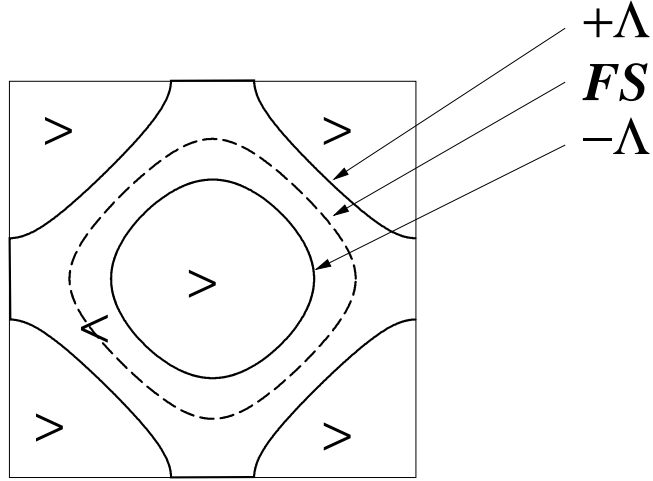,width=11cm}}
\caption{The division of the Brillouin zone into the outer--shell
($>$), the on--shell $(l)$, and the slow $(<)$ modes.}
\label{Fig2}
\end{figure}
We now concentrate to one single step $l\rightarrow l+dl$ of the mode
elimination. We call {\it outer shell} modes the modes already eliminated by
the previous steps (the fast $ (>)$ modes) . The modes inside the shell
$[\Lambda _0 \exp (-l) -\Lambda _0 \exp (-l-dl)]$ are the ones to integrate
out.  We call them {\it on--shell} modes and denote them by $(l)$.  Figure
\ref{Fig2} shows the division of the Brillouin zone into three types of
modes $(>,l$, and $<)$ for the case of the non--half--filled Hubbard model.

If $l$ is not the very first step, the effective action $S_ {\Lambda}$
contains couplings of all orders. Schematically it reads
\begin{equation} \label{action-noeuds}
S_{\Lambda}= S_{0\Lambda}+S_{I\Lambda}=\Gamma _2^{(l)}\bar{\Psi}{\Psi}+\Gamma
_4^{(l)}\bar{\Psi}\bar{\Psi}{\Psi}{\Psi}+ \Gamma _6^{(l)}
\bar{\Psi}\bar{\Psi}\bar{\Psi}{\Psi}{\Psi}{\Psi} + ...
\end{equation}
The summation over all frequencies, momenta, and spins is assumed. The two
point vertex $\Gamma _2$ defines Wick's theorem at step $l$. In particular
the propagator of the ``bare electrons'' $\Gamma _2^{-1}$ changes as we
proceed with the renormalization.

\begin{figure}
\centerline{\epsfig{file=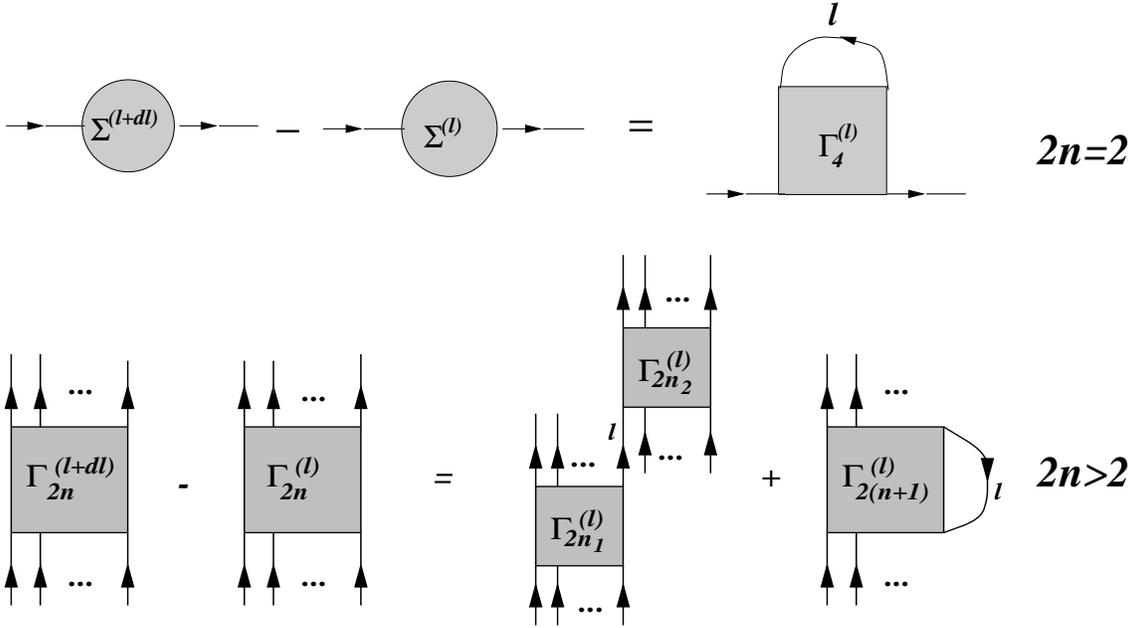,width=15cm}}
\caption{ Polchinski equation for the vertices with $n=2$ and $n>2$ legs.}
\label{Fig3}
\end{figure}

The construction of the effective action one step further (at $l+dl$) is of
the same form as the equation (\ref{LEEA1}), with the difference that now
the on--shell modes play the role of the fast modes.  As we are interested
only in the renormalization of vertices, we can skip the constant $\Omega$
and we get the recursion relation:
\begin{equation} \label{recursion}
S_{\Lambda(l+dl)}=S_{\Lambda(l)}+\delta {S}(l)\;  .
\end{equation}
The contribution $\delta S(l)$ is due to the elimination of $l$ modes. It is
given by the sum of all cumulants made of $S_{I\Lambda} \{ \Psi _l\}$ and
$S_{I\Lambda} \{ \Psi _l,\Psi _< \}$ with two or more legs but now with all
internal momenta constrained to be on--shell.  Now we use the fact that $d
\Lambda$ is infinitesimal: In the expression for $\delta {S}(l)$ only the
terms linear in $dl$ will survive, to make the recursion (\ref{recursion}) a
differential equation for $S_{\Lambda(l)}$.  Generally the cumulants with
$m$ internal lines are proportional to $d\Lambda ^m$ because every internal
line is constraint to the shell.  In principle only diagrams with one
internal line are proportional to $dl$.  If we group the terms with equal
number of legs, we obtain the flow equation for vertices $\Gamma _n^{(l)}$,
known as  Polchinski equation for the vertices
.\cite{Polchinski84,Morris94}
Only two types of diagrams
with one internal line are possible: tree diagrams and loop diagrams.  
The
Polchinski equation for the vertices is shown on fig.\ref{Fig3}.
Its symbolical form for a two--point vertex $(2n=2)$ is 
\begin{equation} \label{Polchinski1}
\frac{\partial}{\partial \Lambda _l}\Gamma_2^{(l)}(K)=-T\sum _{\omega _n'}
\int _{d\Lambda} d^2k' \Gamma _4^{(l)}(K', K, K',K)G_l(K').
\end{equation}
This means that only the loop term renormalizes the selfenergy (see figure
\ref{Fig3}). 
Both loop and tree terms are present in the Polchinski equation for the higher
order vertices:
\begin{eqnarray} \nonumber
\lefteqn{\frac{\partial}{\partial \Lambda _l} \Gamma_{2n}^{(l)}(K_1,...,K_n,
K_{n+1},...,K_{2n})=\sum_{I_1,I_2}T\sum _{\omega _n}
\int _{d\Lambda} d^2k\Gamma_{2n_1}^{(l)}(-K,I_1)G_l(K)
\Gamma_{2n_2}^{(l)}(K,I_2)-} & \\
& - & T\sum _{\omega _n}
\int _{d\Lambda} d^2k \Gamma_{2(n+1)}^{(l)}
(K,K_1,...,K_n,K,K_{n+1},...,K_{2n}) G_l(K) \; .
\label{Polchinski2}
\end{eqnarray}
The two--point vertex defines the  one--particle 
propagator $G_l$ at each step of the renormalization:
 \begin{equation} \label{1pprop}
G_l(K)=(\Gamma _2^{(l)}(K))^{-1}
\end{equation}
We use this {\em renormalized} 
propagator  to construct the Wick theorem.
We name $\Gamma_{2n}^{(l)}(K_1,...,K_n,
K_{n+1},...,K_{2n})$ the vertex with
$2n$ external legs at the step $l$ of the renormalization, with legs 
$\{ K_1,..,K_n\}$ coming in  and 
$\{K_{n+1},...,K_{2n}\}$ coming out.
Symbols $I_1$ and $I_2$ in equation (\ref{Polchinski2}) are disjoint 
subsets $(I_1\cap I_2=\emptyset )$ 
of the energy--momenta such that $I_1 \cup I_2= \{K_1,...,K_{2n}\}$.
The sum runs over all such sets. We have skipped  spin indices for 
simplicity.

We see that the Polchinski flow equation is a functional equation because
all vertices are renormalized as {\it functions} of momenta and frequencies.
It gives the exact renormalization group flow of the model.  The careful
reader will perhaps hesitate at this point: some of the loop diagrams 
 with {\em two} internal lines  as for
example the p-p diagrams are also proportional to just $dl$ if we put the
total momentum and energy to zero.  The same ``anomaly'' happens if the
nesting is perfect for some p-h diagrams. In the usual renormalization group
calculations only these contributions are taken into account\cite{Shankar}
because they give the dominant logarithmic part of the renormalization group
flow.  Then one misses all non--logarithmic or ``almost logarithmic''
physics. This is not a problem in one dimension, for example. But if we are
in two or more dimensions and especially if the nesting is good but not
perfect, it is better to consider the exactly logarithmic configurations of
energy-momentum as exceptions. If we formulate the flow equations correctly
for the general case, the exact logarithmic terms will also appear, as we
will see in the next subsection.

In principle vertices of all orders are created with increasing powers
of the initial coupling $U_0$: It is easy to see that the vertex $\Gamma
_{2n}$, $(n>2)$, is created by the tree term of the Polchinski equation with
power $(n-1)$ of the bare coupling.  This means that the truncation of
the expansion (\ref{action-noeuds}) is equivalent to weak--coupling
perturbation theory.

\subsection{Truncation of the Polchinski equation: one--loop renormalization 
group}
\begin{figure}
\centerline{\epsfig{file=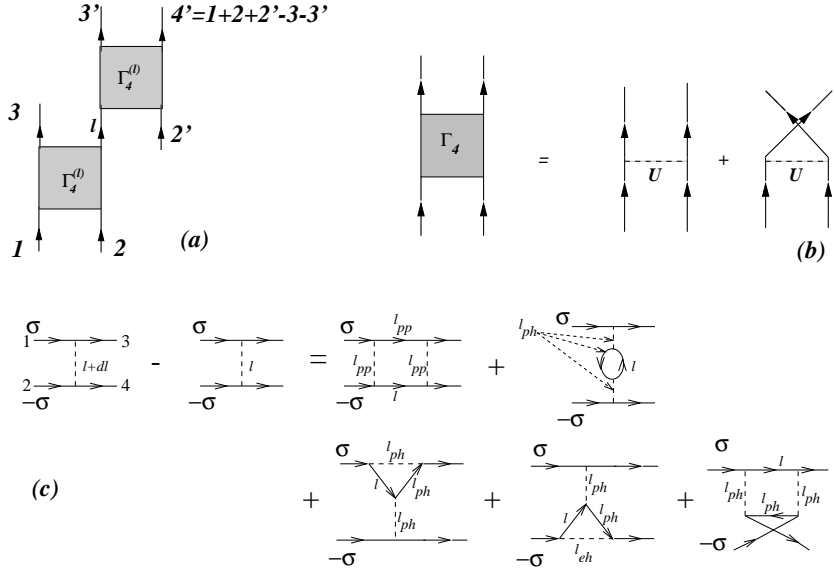,width=11cm}}
\caption{(a) The six--point vertex for one--loop renormalization group.
(b) Relation between the vertex $\Gamma _4$ and the interaction $U$. (c) 
Recursion for the one--loop renormalization of the interaction $U$.}
\label{Fig4}
\end{figure}

The one--loop renormalization for the vertex $\Gamma _4$ (or for the
effective interaction $U_l$) is the perturbative procedure to truncate
the flow equations at order $U^2$.  All terms of order higher than six in the
expansion (\ref{action-noeuds}) are created with a power higher then $2$ of
the interaction by the tree term of the Polchinski equation.  Thus putting
$\Gamma _8=\Gamma _{10}=...=0$, we generate the one--loop renormalization
group.  The only contribution to the vertex $\Gamma _6$ is than the tree
term, made of two $\Gamma _4$ terms connected by one line (see figure
\ref{Fig4}(a)).  The line denoted with $l$ has in principle to be taken
dressed by the selfenergy at the step $l$ defined as
\begin{equation} \label{SFE}
\Sigma _l\equiv \Gamma _2^{(l)} - \Gamma _2^0=\Gamma _2^{(l)} - 
i\omega _n +\xi
_{\bf k} ^0 .
\end{equation}
We assume that the selfenergy remains diagonal upon renormalization.  This
is consistent with the weak--coupling treatment because off--diagonal terms
would imply the existence of some form of long--range order, which is out of
the reach of the present calculation. All we can possible expect from our
calculation is a divergence of some effective interaction signaling the {\em
onset} of long--range order.

We now go back to the formulation in terms of the interaction as defined by
eq.(\ref{act}) and illustrated in fig.\ref{Fig4}(b).  We will skip the spin
indices where they are not necessary.  The differential flow of the six
point function $\Gamma _6$ at step $l$ is according to the fig.\ref{Fig4}(a)
given by
\begin{eqnarray} \nonumber
\lefteqn{d \Gamma _6^{(l)}(K_1,K_2,K_3,K'_2,K'_3) =T \sum_{\omega_n}
 \int_{d\Lambda} d^2k  \; \delta ({\bf k}-{\bf k}_1 
-{\bf k}_2+{\bf k}_3)  \delta_{\omega_n-\omega_{n1}-
\omega_{n2}+\omega_{n3}}} & \\
& \times &  G_l(K) \; U_l(K_1,K_2,K_3) \; U_l(K,K'_2,K'_3)\; . 
\label{gama6}
\end{eqnarray}
The phase space integral is over the shell of thickness $d\Lambda$
corresponding to step $l$, and $G_l(K)$ is the renormalized Green function
at the same step. Physically this is the propagator of an on--shell electron
renormalized by the scattering on the fast electrons.  $U_l(K_1,K_2,K_3)$ is
the effective interaction at  step $l$.  The vertex $\Gamma _6$ at some
step $l$ is  the integral of eq.(\ref{gama6}) over all
steps between $\Lambda =\Lambda_0$ and $\Lambda _l$, that is over all fast
degrees of freedom.  
On the other hand, there is no loop integration in this term: the Dirac 
function in equation (\ref{gama6}) reduces the integral 
$\int _{d\Lambda} d^2k$ to a single point 
${\bf k}={\bf k}_1+{\bf k}_2-{\bf k}_3$ 
$(={\bf k}_{3'}+{\bf k}_{4'}-{\bf k}_{2'})$.
To get  $\Gamma _6$  we can thus skip the integration over $dl$ and take just 
care of  momentum conservation.
The effective action at the step $l$ then reads
\begin{eqnarray} \nonumber
\lefteqn{S_l=\sum _{\omega _n,{\bf k}}\bar{\Psi} _{K, \sigma}
(i\omega _n-\xi_{\bf k} + 
\Sigma _l(K)) {\Psi _{K, \sigma}}  \Theta (\Lambda (l)-|\xi_{\bf k}|)}& & \\
\nonumber
& +& \frac{1}{2}T^{3}\sum _{\sigma \sigma '}
\sum_{1,2,3}
U_l(K_1,K_2,K_3)\Theta ^{(\Lambda (l))}_{\bf
k_1,k_2,k_3,k_4}
\bar{\Psi}_{\sigma K_3}\bar{\Psi}_{\sigma 'K_4}{\Psi}_{\sigma 'K_2}
{\Psi}_{\sigma K_1} \\
\nonumber
& + & T^{5} 
\sum _{\sigma ,\sigma ', \sigma ''}
\sum _{1,2,3,2',3'} 
[ \Theta ^{(\Lambda (l))}_{\bf k_1,k_2,k_3,k_2',k_3',k_4'} 
\Theta (|\xi_{\bf k}|-\Lambda (l)) \\
 \label{action1} 
& & \times G_{l'}(K) U_{l'}(K_1,K_2,K_3) U_{l'}(K,K_2',K_3')
\bar{\Psi}_{\sigma K_3}\bar{\Psi}_{\sigma' K_{3}'}\bar{\Psi}_{\sigma ''K_{4'}}
{\Psi}_{\sigma ''K_{2'}}{\Psi}_{\sigma 'K_2}{\Psi}_{\sigma K_1} ],
\end{eqnarray}
where $l'=\ln \Lambda _0/|\xi_{\bf k}|$ (i. e. $\xi _{\bf k}=\Lambda(l')$)
is the scale fixed by external momenta, $K=K_1+K_2-K_3$ and the
energy--momentum $4'=1+2+2'-3-3'$. The summations over 1,2,3,... run over 
corresponding Matsubara frequencies and momenta.
The term of the sixth order contains the
interactions and Green functions from  former steps $l'<l$ of the mode
elimination since 
only fast degrees of freedom contribute to $\Gamma _6^{l}$ so that
$l'\leq l$. This constraint is imposed to the sextic term 
of action (\ref{action1}) by
$\Theta (|\xi_{\bf k}|-\Lambda (l))$.
The
functions  $\Theta ^{(\Lambda )}_{\bf k_1,k_2,k_3,k_4}$ and 
$\Theta ^{(\Lambda )}_{\bf k_1,k_2,k_3,k_2',k_3',k_4'} $ constrain
the  momenta in arguments to be slow modes 
(inside a shell of thickness $\pm \Lambda$
around the Fermi surface). 
This simply means that the fields described by the effective action at cutoff
$\Lambda (l)$ are inside  the cutoff range.

\begin{figure}
\centerline{\epsfig{file=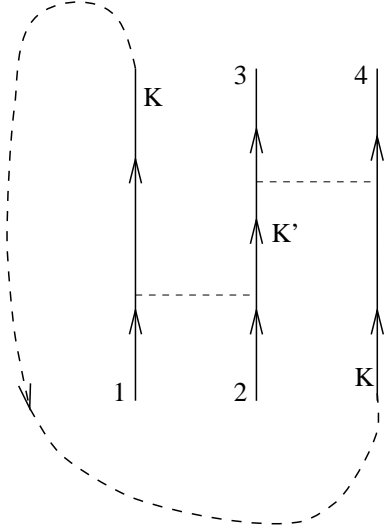,width=5cm}}
\caption{ Construction of the p-p diagram 
from the six--leg vertex.}
\label{Fig4bis}
\end{figure}

If the initial interaction $U_0$ is spin independent, the renormalized
interaction $U_l$ will remain spin independent as well. It is thus not
necessary to worry about spin indices and all two--particle interactions are
given only by {\it one} function $U_l(1,2,3)$. The detailed justification for
that is given in the appendix \ref{inv_interactions}.  The differential flow
of $U_l(1,2,3)$ is readily obtained  applying the loop term of the
Polchinski equation  (the second term in  equation (\ref{Polchinski2}) 
and in  figure
\ref{Fig3}(b))
 to the six--leg part of the effective
action (\ref{action1}). 
At first sight, two kinds of diagrams are created: one--particle 
reducible (1PR) and one--particle 
irreducible  (1PI) ones. 
We will show that only 1PI diagrams contribute to the renormalization 
of the effective interaction :
We can try to construct the 1PR diagram by contracting 
legs 2' and 4' in  figure \ref{Fig4}(a). This immediately implies that the
internal line denoted with $l$ and the leg 3' carry the same momentum. 
This momentum corresponds to some fast mode since  line $l$ is already 
integrated out. The conclusion is that the resulting four--point 1PR vertex
can not be a vertex of the effective action (\ref{action1}) since
this action contains only slow modes. Consequently, only 1PI diagrams 
renormalize the effective interaction between slow electrons.
The resulting diagrams are all topologically different two--particle loops as
shown on figure
\ref{Fig4}(c).
The first diagram is a p-p diagram and
the others are p-h diagrams.  
Let us ilustrate how we obtain the first diagram in fig. \ref{Fig4}(c).
The procedure is shown on fig.\ref{Fig4bis}.
The diagram represents the p-p contribution to the effective interaction
$U_l(1,2,3)$ due to the elimination of the infinitesimal shell at 
step $l$.
We take the six--leg diagram with the configuration of 
external momenta shown in the figure and with  legs $K$ being on--shell.
Their contraction (dashed line) is done precisely at  step $l$.
The contraction $K'$ was done at a previous step $l_{pp}$  fixed by 
momenta $K$, $K_1$, and $K_2$ (see equation (\ref{gama6})).
$K_1+K_2=Q_{pp}\equiv (\omega _{npp},{\bf q}_{pp})$ is the total 
energy--momentum in the p-p process. The scale $l_{pp}$ is then given by
\begin{equation} \label{lpp}
l_{pp}=-\ln \frac{\xi _{{\bf k-q}_{pp}}}{\Lambda _0}\; .
\end{equation}
Similar constructions give all other (p-h) diagrams. One has to take care of 
both direct and exchange interactions (fig. \ref{Fig4}(b)) to get four 
different graphs. The interactions and one--particle propagators are to be 
taken at the scale $l_{ph}\leq l$. This scale is determined by the momentum 
transfer ${\bf q}_{ph}$ as
\begin{equation} \label{lph}
l_{ph}=-\ln \frac{\xi _{{\bf k+q}_{ph}}}{\Lambda _0}\; .
\end{equation}
In second, third, and fourth diagram in fig. \ref{Fig4}(c) 
the energy--momentum 
transfer is $Q_{ph}=K_1-K_3$, while in the last diagram $Q_{ph}=K_1-K_4$.

We see that even though the
Polchinski equation appears as local in $l$, the flow at step $l$ depends on
the Green functions and interactions at the former steps $l_{pp},l_{ph}\leq
l$. 
The reason for that is the
dependence of the six point function on 
two--particle interaction and on one--particle propagator at 
all former steps.

In fig.\ref{Fig4}(c) 
the internal lines labeled with $l$ are to be integrated over two
infinitesimal shells of the width $d \Lambda $ at $\xi _0 =\pm \Lambda_l$.
For this purpose we pass from the Cartesian measure $dk_xdk_y$ to the
measure
\begin{equation} \label{measure}
 \int _{\xi}^{\xi+d\Lambda}d\xi' \oint
\frac{ds}{v(s,\xi ')} \;, 
\end{equation}
where $s$ are lines (or surfaces in $D>2$) of constant energy $\xi ({\bf
k})$ and ${v(s,\xi ')}$ is the group velocity as defined by $\partial\xi
({\bf k})/\partial k_{\perp}$ ($k_{\perp}$ is the component of the momentum
perpendicular to the equal--energy lines.
We will use  measure (\ref{measure}) in what follows, where we write
the analytic expression
for the flow of function $U_l(1,2,3)$.

From the diagrams in fig.\ref{Fig4}(c) one obtains the following expression
\begin{eqnarray} \nonumber
\lefteqn{\frac{\partial{U_l}}{\partial{l}}=\beta (l,\{ U \} )} & & \\
\label{Beta.fctn}
& = & \beta _{pp}\{ U,
U\} +2{\beta}_{ph}\{ U,U\}-
{\beta}_{ph}\{ U,XU\}-{\beta}_{ph}\{ XU,U\}-X{\beta}_{ph}\{ XU,XU\}.
\end{eqnarray}
$\beta$ is a four point object and a bilinear functional of $U_{l'}$,
$(l'\leq l)$. The operator $X$ is the exchange operator acting on a
four--point function: $XF(1,2,3,4)\equiv F(2,1,3,4)$.  $\beta _{pp}$ and
${\beta}_{ph}$ are the p-p and p-h parts of the $\beta$ function given by
\begin{eqnarray} \label{Betaee}
\beta _{pp}\{ U,U\} & = & \left( \Xi \{ U,U\} +\Xi \{ XU,XU\} \right)\\
\label{Betaeh}
\beta _{ph}\{ U_1,U_2\} & =& \left( \Pi \{ U_1,U_2\} +{\cal T}
\Pi \{ U_1,U_2\}\right)
\; .
\end{eqnarray}
${\cal T}$ is the time inversion operator acting on a four point function:
${\cal T}F(1,2,3,4)\equiv F(3,4,1,2)$.  
The functions $\Xi$ and $\Pi$
correspond to the on--shell integrals of the p-p and p-h bubbles:
\begin{eqnarray} \nonumber
\lefteqn{\Xi \{ U,U\}(K_1,K_2,K_3,K_4)} & & \\
\nonumber 
& = & \frac{-\Lambda _l}{(2\pi)^2}\sum _{\nu =+,-}
\int \frac{ds_{\nu}}{v_{\nu}}
\Theta
\left(
 |\xi _{{\bf k}_{\nu}-{\bf q}_{pp}}|-\Lambda _l\right) 
T\sum _{\omega _n}\;
G_l(K_{(\nu)})\; G_{l_{pp}}(-K_{(\nu)}+Q_{pp})\\
\label{Xi}
& & \times  U_{l_{pp}}(K_1,K_2,K_{(\nu )})U_{l_{pp}}(K_3,K_4,K_{(\nu )})\; ,
\end{eqnarray}
\begin{eqnarray} \nonumber
\lefteqn{\Pi \{ U_1,U_2\}(K_1,K_2,K_3,K_4)} & & \\
\nonumber
& = & \frac{-\Lambda _l}{(2\pi)^2}\sum _{\nu =+,-}
\int \frac{ds_{\nu}}{v_{\nu}}
\Theta
\left( |\xi _{{\bf k}_{\nu}+{\bf q}_{ph}}| -\Lambda _l\right) 
T\sum _{\omega _n}\;
G_l(K_{(\nu)})\; G_{l_{ph}}(K_{(\nu)}-Q_{ph}) \; . \\
\label{Pi}
& & \times U_{1,l_{ph}}(K_1,K_{(\nu )},K_3)U_{2,l_{ph}}(K_4,K_{(\nu )},K_2)\; .
\end{eqnarray}
$U_1$ et $U_2$ can be $U$ or $XU$ as required by eq.(\ref{Beta.fctn}).  The
summation over index $\nu=+,-$ is over two shells at $\xi _0=\pm \Lambda_l$;
the velocities are $v_{\nu}=v(s_{\nu},\xi =\nu \Lambda)$ and $K_{\nu}$
symbolizes $({\bf k}_{\nu},\omega _n)$.  The quantity
$$
Q_{pp}=(\omega _{n,pp},{\bf q}_{pp})=
K_1+K_2
$$
is the total energy--momentum and 
$$Q_{ph}=(\omega _{n,ph},{\bf
q}_{ph})=K_1-K_3,
$$ 
is the energy--momentum transfer between the currents (1,3) and (2,4) where
 1,2,3 and 4 are the external variables of $\Xi$ and $\Pi$.  The scales
 $l_{pp}$ and $l_{ph}$ are defined by expressions (\ref{lpp}) and (\ref{lph}).
As already discussed, $l_{pp}$ and $l_{ph}$ depend on the integration 
variable ${\bf k}_{\nu}$ and
on the configuration of the external energies--momenta.  
Let's call the
external legs of the total $\beta$ function (\ref{Beta.fctn})
$\tilde{1},\tilde{2},\tilde{3}$ and $\tilde{4}$.  Note that the operator $X$
exchanges the external legs $\tilde{1}$ and $\tilde{2}$ in the last term of
this expression. This means that the energy--momentum transfer in this term
is $Q_{ph}=K_{\tilde{1}}- K_{\tilde{4}}$ and not
$K_{\tilde{1}}-K_{\tilde{3}}$ as in the first three e-h terms. In the
standard language (see for example ref.\onlinecite{Shankar}) the p-h terms
with transfer $1-3$ are called zero--sound (ZS) terms, and the terms with
transfer $1-4$ are ZS'.

Let us explain briefly how we  obtained the flow equation
(\ref{Beta.fctn}). The first term is simply the p-p loop with 
$U$ interaction. The remaining terms are different versions of the p-h loops, 
corresponding respectively to the p-h diagrams in the figure \ref{Fig4}(c).
They can all be seen as a single loop $\beta_{ph}$
(with the topology of the second diagram in fig.\ref{Fig4}(c)), given by  
(\ref{Betaeh}) and (\ref{Pi})
by performing 
appropriately the 
operation $X$. The third and the fourth 
graph can be drawn as the second one by one exchange:
In the third term we replace the upper interaction line $U$ by $XU$ and in the 
fourth the lower one. After this manipulation both diagrams look like
the second diagram.
The last graph is 
more complicated: one has to perform $X$ upon both interactions and upon the 
whole graph to see it as $\beta_{ph}$.
The factor 2 before the second term is due to 
spin summation in the loop. All other diagrams have fixed spin.

The flow of the effective action (\ref{action1}) is still not completely
determined because we do not know how the self--energy $\Sigma _l(K)$ is
renormalized. The differential flow for $\Sigma _l(K)$ is readily found from
the Polchinski equation for the two--point function 
(\ref{Polchinski1}) shown graphically in fig.\ref{Fig3}. 
In
the language of the effective interaction $U_l(1,2,3)$ this gives Hartree
and Fock like contributions shown in the fig.\ref{Fig4}(c).  We get the
renormalization equation
\begin{equation} \label{alpha.fctn}
\frac{\partial \Sigma}{\partial l(K_1)}=\alpha _{\text{Hartree}}\{ U _l\}(K_1) 
+ \alpha_{\text{Fock}}\{ U_l \}(K_1) + \alpha_{\mu}^{\text{hom.}}(l).
\end{equation}
The first term is the Hartree term
\begin{equation} \label{hom.}
\alpha_{\text{Hartree}}\{ U _l\} (K_1) =\frac{\Lambda}{(2\pi)^2}\sum _{\nu
=+,-} \int \frac{ds_{\nu}}{v_{\nu}} T\sum _{\omega _n} G_l(K_{\nu})
\frac{1}{2}(3-X)U_l(K_1,K_{\nu},K_1)
\end{equation}
and the second is the exchange term
\begin{equation} \label{ex}
\alpha _{\text{Fock}}\{ U _l\} (K_1) =-\frac{\Lambda}{(2\pi)^2}\sum _{\nu =+,-}
\int \frac{ds_{\nu}}{v_{\nu}} T\sum _{\omega _n} G_l(K_{\nu})
\frac{1}{2}(1-X)U_l(K_{\nu},K_1,K_1)\; .
\end{equation}
The third term is added to cancel the  chemical potential renormalization due 
to the homogeneous part of the direct term
\begin{equation} \label{hom}
\alpha _{\mu}^{\text{hom.}}(l)
=-\frac{1}{(2\pi)^2} \int d^2 k_1 \; \beta ^{-1}\sum _{\omega _{n1}} 
\alpha _{\text{Hartree}}\{ U _l\} (K_1)\; .
\end{equation}
If the initial interaction has no dynamics, the first nontrivial
contributions to the flow of the selfenergy comes from the {\it
renormalized} and not the bare interaction.  As the interaction is
renormalized by one--loop processes this implies that the interesting part
of the flow of $\Sigma _l(K)$ is given by two loops.  On the level of the
present one--loop calculation it is thus consistent to neglect selfenergy
corrections.  This is what we do in the subsequent one--loop renormalization
of the Hubbard model.

\section{Renormalization group for the Hubbard model}

In this section we will apply  the above renormalization--group procedure 
 to the Hubbard model.
The model  is given by equations (\ref{Hubbarddir}), (\ref{Hubbard}), 
and (\ref{dispersion}).
The initial action $S$ for the Hubbard model is given by  expression 
(\ref{act}),  with the dispersion (\ref{dispersion}) and with the initial
 interaction $U_0=\mbox{constant} (K_1,K_2,K_3)$. 
The interaction will depend more and more on $(K_1,K_2,K_3)$ 
as we go on with the renormalization (as $l$ increases) so that we will
see  at work the functional aspect of our renormalization group.
We will complete our analysis by the renormalization of  two--particle
correlation functions.

\subsection{Renormalization of the interaction}

\begin{figure}
\centerline{\epsfig{file=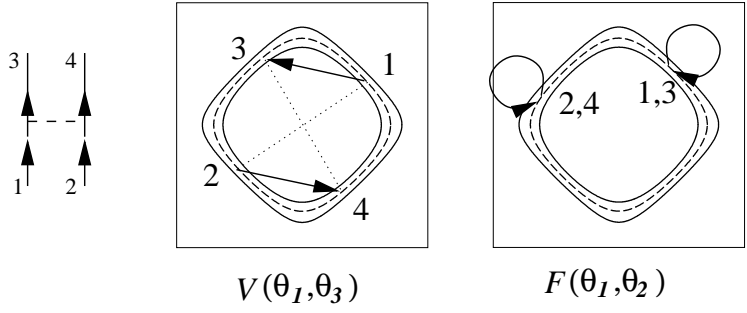,width=11cm}}
\caption{The marginal interactions in the BCS regime.}
\label{Fig5}
\end{figure}

If we neglect the selfenergy corrections, the flow of the effective 
interaction is completely determined by the expressions 
(\ref{Beta.fctn}-\ref{Pi}) with the bare propagators instead of 
the renormalized ones:
\begin{equation} \label{prop}
G_l(K)\rightarrow G_0(K)\equiv (i\omega _n-\xi_{\bf k})^{-1}
\end{equation}
The effective interaction is a function of three energy-momenta. 
This makes formulae (\ref{Xi}) and (\ref{Pi}) very complicated.
For that reason we will consider only the marginal part of the dependence
of $U_l$ on energy-momenta. This approximation is justified by the zero--order 
scaling and power--counting arguments. \cite{ZS_prb,ZS_zfp,Shankar}
For example, in one dimension this procedure justifies the well--known g--ology
model\cite{Rev1}: one adds an index $i$ to the electrons so that all 
electrons moving to the left have $i=-$ and all right--movers have $i=+$. 
Then the marginal interactions do not depend on impuls $k$ and energy 
$\omega$  of 
the electrons in interaction, but only on their indices $i$. 
This can be seen as parametrization
 of the interactions as if the electrons were on  the 
Fermi surface (or points in 1D) with $\omega=0$.
In two dimensions  the marginal 
interactions depend only on polar coordinates of the wave vectors. 
Only the interactions between electrons at the Fermi 
surface are then kept and, if the Fermi surface is not nested, 
one gets the  LEEA for the Fermi liquid.
\cite{ZS_zfp,Shankar}. The marginal processes in that case are
\begin{equation} \label{V-marginal}
V(\theta _1,\theta _2)= U_l(K_1,-K_1,K_2) \; ; \omega _{1,2}=0 \; ;
\; \xi_{{\bf k}_{1,2}}=0
\end{equation}
and
\begin{equation} \label{F-marginal}
F(\theta _1,\theta _2)= U_l(K_1,K_2,K_1) \; ; \omega _{1,2}=0 \; ;
\; \xi_{{\bf k}_{1,2}}=0.
\end{equation}
$V$ is the pairing amplitude and $F$ is the forward scattering related to 
the Fermi liquid parameter. Both kinds of processes are shown in  figure 
\ref{Fig5} for the case of the Hubbard model far from the half filling. 
We have analyzed in detail this problem in a former article.\cite{ZS_prb}
\begin{figure}
\centerline{\epsfig{file=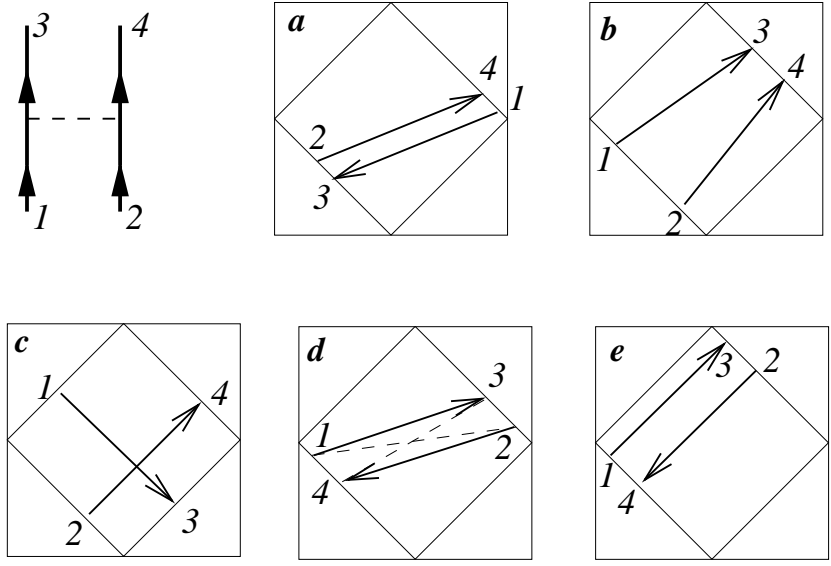,width=11cm}}
\caption{Some of the marginal processes for the square Fermi surface.}
\label{Fig6}
\end{figure}
Let us now concentrate on the square Fermi surface, for the half filled 
Hubbard model. The processes between electrons on the Fermi surface are now 
labeled with three variables instead of  two as in the case of 
the Fermi surface without nesting : if we put  particles 1, 2 and 3 
{\it anywhere} on one side or on two opposite sides of the square, the fourth 
falls exactly on the square as well. This is due to  the perfect 
flatness of the Fermi surface and to the marginality of the umklapp processes.
A few examples of marginal interactions between the electrons on the square
are shown on  figure \ref{Fig6}. The interaction depends only on the 
positions of the particles on the square.
The ``angle'' $\theta$ can be defined in a way shown on figure \ref{Fig7}.
\begin{figure}
\centerline{\epsfig{file=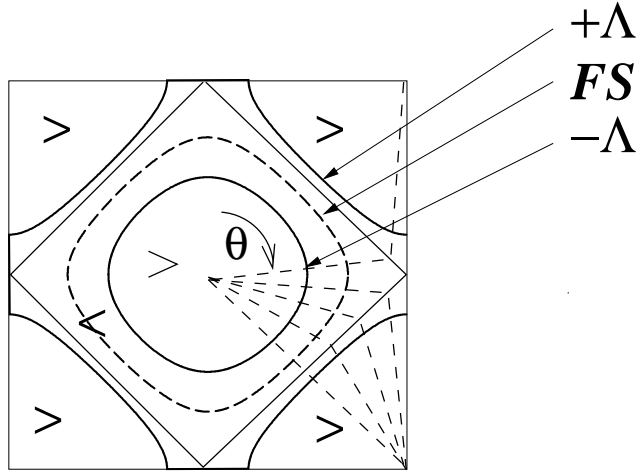,width=11cm}}
\caption{ The
organization of the mode elimination.  Dashed
lines are the lines of constant ``angle'' $\theta$.}
\label{Fig7}
\end{figure}
It is important to notice that even if the filling is not exactly one--half 
(and the Fermi surface not exactly square), all above 
interactions will still be 
important, as long as the effective phase space is open i.e. when 
$\Lambda >|\mu|$
(as on figure \ref{Fig7}). 
We thus take  as marginal 
all effective interactions viewed as functions of three angles $\theta$ of the 
particles:
\begin{equation} \label{U-marginal}
U_l(K_1,K_2,K_3)\rightarrow U_l(\theta _1,\theta _2,\theta _3) \; ; 
\omega _{1,2,3,4}=0 \; ;
\; {1,2,3,4}\mbox{ are on the square}.
\end{equation}
When the cutoff becomes smaller than the 
chemical potential, we are back to the non nested case, in which the functions
$V$ and $F$ are the 
marginal interactions. 
They read 
\begin{equation} \label{V-marginal1}
V_l(\theta _1,\theta _2)= U_l(\theta _1,\theta _1+\pi,\theta _2)\; ,
\hspace{10mm}F_l(\theta _1,\theta _2)= U_l(\theta _1,\theta _2,\theta _1)\; .
\end{equation}
Altogether, for the half and almost half filled Hubbard model the 
function $U_l(\theta _1,\theta _2,\theta _3)$  given by (\ref{U-marginal}) 
contains all  marginal  scattering processes. 
The renormalization group analysis is now much  simpler  
because we deal with
a function of three  variables instead of six.

We will now derive the flow equation for $U_l(\theta _1,\theta _2,\theta _3)$ 
at zero temperature. The pleasant aspect of the Kadanoff--Wilson--Polchinski
mode elimination technique 
at $T=0$ is that $\Lambda$ can then be interpreted as the 
temperature. Namely, the interaction at some temperature
$T$ is renormalized mainly by virtual processes involving ``quantum'' 
electrons, those with
energy larger than $T$, having almost the distribution of the $T=0$ electrons.
This is exactly what we do with the renormalization group:  only 
modes with $|\xi _{\bf k}| > \Lambda$ 
are involved in the virtual processes renormalizing $U_l$.
Consequently,  $\Lambda$ is not only the 
measure of how many electrons are already integrated out :  it has a 
{\em physical}
meaning of the effective temperature.

Replacing $G_l(K)$ by $G_0(K)$ and 
$U_l(K_1,K_2,K_3)$ by $U_l(\theta _1,\theta _2,\theta _3)$  in 
expressions (\ref{Xi}) and (\ref{Pi}), we can get interactions out of
Matsubara summations and 
perform the summations
analytically. After taking the $T\rightarrow 0$ limit we obtain
\begin{eqnarray} \nonumber
\lefteqn{\Xi \{ U,U\}(\theta _1,\theta _2,\theta _3)
 =\frac{-2}{(2\pi)^2}\sum _{\nu =+,-}
\int d\theta {\cal J}(\nu\Lambda,\theta)
\frac{
\Theta
\left(\nu\xi _{{\bf k}_{\nu}-{\bf q}_{pp}}
 \right) 
\Theta
\left(|\xi _{{\bf k}_{\nu}-{\bf q}_{pp}}|-\Lambda
 \right) 
}{1+\frac{\nu}{\Lambda}\xi _{{\bf k}_{\nu}-{\bf q}_{pp}}}
\times } & \\
& \times &   U_{l_{pp}}(\theta_1,\theta_2,\theta)
U_{l_{pp}}(\theta_3,\theta_4,\theta)\; ,
\label{Xi-theta}
\end{eqnarray}
\begin{eqnarray} \nonumber
\lefteqn{\Pi \{ U_1,U_2\}(\theta _1,\theta _2,\theta _3)
 =\frac{2}{(2\pi)^2}\sum _{\nu =+,-}
\int d\theta {\cal J}({\nu}\Lambda,\theta)
\frac{\Theta
\left(-\nu\xi _{{\bf k}_{\nu}+{\bf q}_{ph}}
 \right) 
\Theta
\left(|\xi _{{\bf k}_{\nu}+{\bf q}_{ph}}|-\Lambda
 \right) }
{1-\frac{\nu}{\Lambda}\xi _{{\bf k}_{\nu}+{\bf q}_{ph}}}
\times  } & \\
& \times &   U_{1,l_{ph}}
(\theta_1,\theta,\theta_3)
U_{2,l_{ph}}(\theta_4,\theta,\theta_2)\; .
\label{Pi-theta}
\end{eqnarray}
${\bf k}_{\nu}$ is the momentum of a particle at the angle $\theta$ with energy
$\xi =\nu \Lambda$.  ${\cal J}(\epsilon,\theta)\equiv 
J[(x,y)/(\epsilon,\theta)]=\partial s/\partial \theta /v(\theta,\epsilon)$ 
is the Jacobian of the transformation from
rectangular coordinates in momentum space to  polar coordinates.
One should not forget that the scales $l_{pp}$ and $l_{ph}$, which make the 
flow equation non--local, depend on external momenta and on the integration
variable $\theta$ through ${\bf q}_{pp}$, ${\bf q}_{ph}$ and 
${\bf k}_{\nu}$ as given by the relations (\ref{lpp}, \ref{lph}).
$U_1$ and $U_2$ represent $U$ or $XU$ as required by  equation 
(\ref{Beta.fctn}).
Equation (\ref{Xi-theta}) gives the  leading logarithmic flow 
in the p-p channel for the
configuration of momenta with ${\bf q}_{pp}=0$ while equation 
(\ref{Pi-theta}) gives the leading logarihmic flow in the p-h channel 
only exactly at 
half filling for ${\bf q}_{ph}=(\pi, \pi)$.
In  the standard renormalization-group procedure \cite{Shankar}, only these
configurations are taken into account.
We see that in our formalism they are taken into account on equal 
footing with all other scattering processes, 
with any values of ${\bf q}_{pp}$ and
${\bf q}_{ph}$, as illustratd on figure \ref{Fig6}. 
The processes with the leading logarithmic renormalization in one channel
and with less strong but still important flow in the other channel (as the
processes in  figures \ref{Fig6}(d) and (e)) are the processes which couple
strongly both renormalization channels.
For exemple, the process in figure \ref{Fig6}(d) has leading logarithmic
renormalization in p-p channel and less strong 
(but still logarithmic!, because of partial nesting) 
renormalization in p-h channel, 
while the process in figure \ref{Fig6}(e) has perfect nesting and, 
consequently,
leading lorarithm in p-h channel and weaker logarithmic flow in p-p channel.

\begin{figure}
\centerline{\epsfig{file=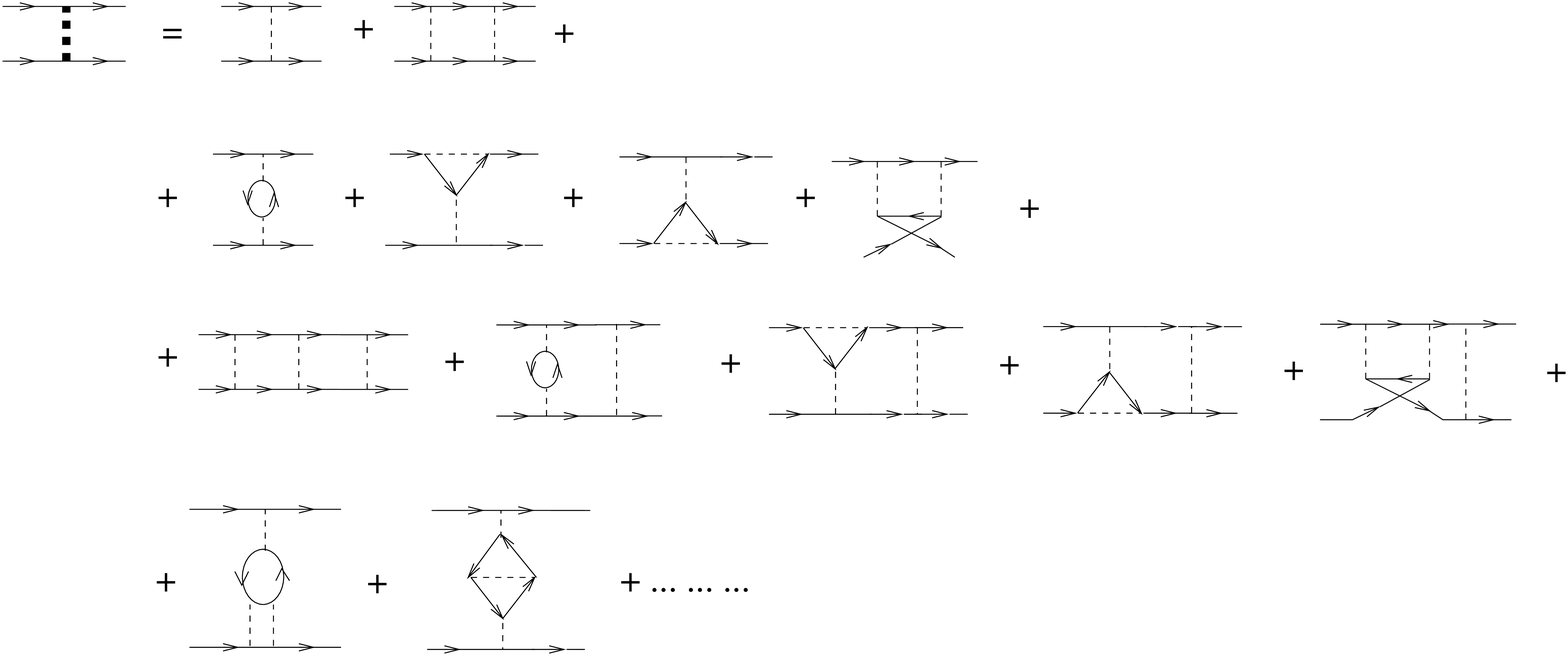,width=11cm}}
\caption{The parquet summation.}
\label{Fig8}
\end{figure}

If we want to see which series of diagrams is generated by our renormalization
group we have to solve the differential equation 
(\ref{Beta.fctn}) for $U_l$ in iterations  of the bare interaction $U_0$. 
The obtained series  is exactly the parquet summation. 
It is constructed from all iterations of five basic loop diagrams from 
figure \ref{Fig4}(c).
A few lowest order parquet diagrams are shown on  figure \ref{Fig8}. 

\begin{figure}
\centerline{\epsfig{file=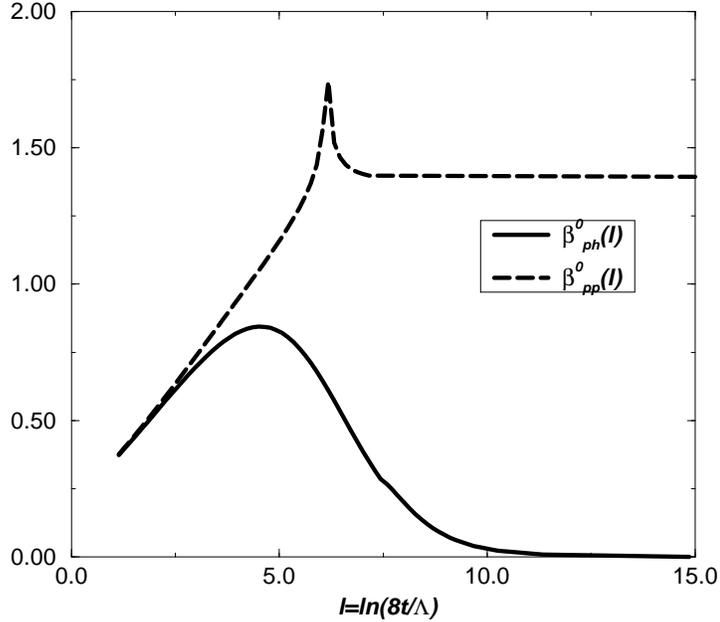,width=11cm}}
\caption{The quantities $\beta _{pp}^0(l)$ and $\beta _{ph}^0(l)$. 
The crossover is at $l=l_{\mu}=6$}
\label{Fig9}
\end{figure}

An important aspect of the non half filled Hubbard model is that it can not be
solved by a scale invariant renormalization group. 
The finite chemical potential determines the intrinsic scale, 
the physical interpretation of which  is 
 the crossover between two different renormalization
regimes. The crossover can be seen from the {\it explicit} scale 
dependence of p-p and p-h differential loops $\Xi(l)$ and $\Pi(l)$.
We can define the quantities 
\begin{equation} \label{tend-ee}
\beta _{pp}^0(l) = \Xi _l\{ 1,1 \}
_{q_{pp}=0},
\end{equation}
\begin{equation} \label{tend-eh}
\beta _{ph}^0(l)= \Pi _l\{ 1,1 \}
_{\bf{q}_{ph}=(\pi,\pi)} \; .
\end{equation}
They measure respectively  the dominant parts of 
p-p and p-h renormalization tendencies. 
      The 
configurations of momenta are chosen to give the most important flow :
for the p-p channel at zero total momentum and for the p-h channel         
at the antiferromagnetic wave vector.
The quantities $\beta _{pp}^0(l)$ and $\beta _{ph}^0(l)$ are shown in 
the figure \ref{Fig9} for  finite chemical potential 
$\mu =-\Lambda _0 exp(-l_{\mu})$. 
We see that for $l<l_{\mu}$ both 
differential loops have linear dependence in logarithmic variable $l$; the 
total (integrated)
 loops are thus square logarithmic, as is known for the half filled band.
When $l>l_{\mu}$ the function $\beta _{pp}^0(l)$ crosses over to  constant
which gives the logarithm of the Cooper bubble. 
$\beta _{ph}^0(l)$ decays exponentially as $\exp (-2l)\sim \Lambda ^2$: 
the nesting does not exist any more and the p-h flow crosses over to 
irrelevance. We call the first regime the 
parquet regime because both loops are 
important. The second regime, in which  only the Cooper channel flows,
 we call BCS regime.
The topology of the effective 
phase space in the parquet regime is open (see figure 
\ref{Fig7}) and, in BCS regime, the phase space is a regular closed ring 
around the Fermi surface, as on figure \ref{Fig5}.
The peak of $\beta _{pp}^0(l)$ at $l=l_{\mu}$ is the enhancement due to
 van Hove singularity. The peak does not exist in $\beta _{ph}^0(l)$ because of
the $\Theta$--function constraint in (\ref{Pi-theta}).
As we will see later, the renormalization in the parquet regime will give 
rise to  precursors of a strong coupling fixed point with dominant 
antiferromagnetic correlations while in the BCS regime only a Cooper--like 
instability is possible.

It is difficult to read  from the sole flow of the interaction $U_l$, what 
kind of correlations are enhanced and possibly divergent. For that purpose
we have to calculate the renormalization of 
the correlation functions.

\subsection{Renormalization of the correlation functions}

\begin{figure}
\centerline{\epsfig{file=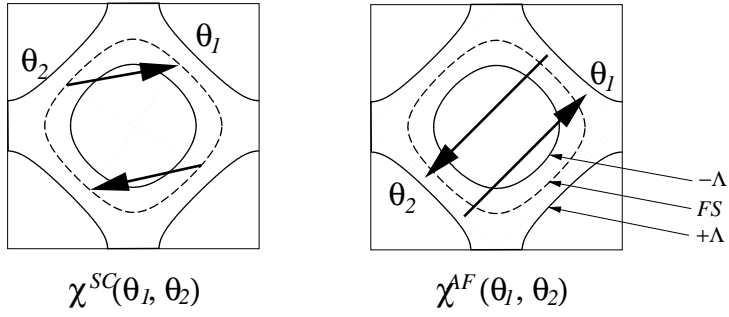,width=11cm}}
\caption{The angle--dependent correlation functions.}
\label{Fig10}
\end{figure}

It is well known \cite{ZS_prb,ZS_zfp,Shankar,Weinberg} that, in studying the
anisotropic superconductivity, one has to consider the pairing 
amplitude as a function of two angles, $V(\theta _1,\theta _2)$. The angles 
determine the angular positions of the Cooper pairs annihilated 
$(\theta _1)$ and created $(\theta _2)$ in the scattering. 
This interaction will be intimately related to the superconducting 
correlation function $\chi ^{SC}(\theta _1,\theta_2 )$. In the same spirit 
we can define the correlation function for the antiferromagnetism dependent on 
two angles. 
We will define both correlation functions in the following way
\begin{eqnarray} \nonumber
\lefteqn{\chi _{\bf q}^{\delta}(\theta_1,\theta_2;|\tau _1-\tau_2|)} & & \\
& = &
\int _>d\epsilon _1 \int _>d \epsilon_2 \; {\cal J}(\epsilon_1,\theta_1)
 {\cal J}(\epsilon_2,\theta_2) 
\langle \hat{\Delta}_{\bf q}^{\delta}(\epsilon_1,\theta_1; \tau_1)
\bar{\hat{\Delta}}_{\bf q}^{\delta}(\epsilon_2,\theta_2; \tau_2)
\rangle,
\label{Correl-funct}
\end{eqnarray}
with $\delta=SC,AF$ (``superconductivity'' or ``antiferromagnetism'').
 The symbols ``$>$'' mean that the energy integrations run 
over energies {\it outside} of the shell $\pm \Lambda$.
Consequently, $\chi ^{SC}$ and $\chi ^{AF}$ are  interpreted as  the 
susceptibilities at the temperature $T=\Lambda$. They measure the response 
of  outer shell electrons for given $\Lambda$.
The order parameter variables are
\begin{equation} \label{SC-def}
\hat{\Delta}_{\bf q}^{SC}(\epsilon,\theta; \tau)\equiv \sum
_{\sigma}\sigma\Psi_{\sigma,{\bf k}}(\tau)
\Psi_{-\sigma,-{\bf k}+{\bf q}}(\tau),
\end{equation}
\begin{equation} \label{SDW-def}
\hat{\Delta}_{\bf q}^{AF}(\epsilon,\theta; \tau)\equiv \sum
_{\sigma}\bar{\Psi}_{\sigma,{\bf k}}(\tau)
\Psi_{-\sigma,{\bf k}+(\pi,\pi)+{\bf q}}(\tau),
\end{equation}
where {\bf k} is given by the angle $\theta$ and the energy $\epsilon$.
The figure \ref{Fig10} illustrates what configurations of four angles are 
described by the correlation functions $\chi ^{SC}(\theta _1,\theta_2 )$ and
$\chi ^{AF}(\theta _1,\theta_2 )$: the first measures the correlation of one 
cooper pair at $\theta _1$ with the other at $\theta_2$ and the second
represents the correlation of the momentum $(\pi, \pi)$ p-h pair
at   $\theta _1$ with the other p-h pair at $\theta _2$. 
The correlation functions can be seen as response functions of the system to 
an infinitesimal external field, as  was done  by Bourbonnais and Caron
\cite{Bourbonnais} in one dimension. We will generalize this procedure to
two dimensions. We  add to the action $S_{\l=0}$ the term
\begin{equation} \label{ext-fields}
S\{ h\} _{l=0}=\int d\tau \int d{\bf q} \int d\theta
\left[
\int d\epsilon {\cal J} (\epsilon,\theta) \hat{\Delta}_{\bf
q}^{\delta}(\epsilon,\theta;\tau)
\right] \bar{h}_{\bf
q}^{\delta}(\theta;\tau) +\mbox{h.c.}
\end{equation}
The external  angle dependent fields 
$\bar{h}_{\bf q}^{\delta}(\theta;\tau)$ are the source fields 
coupled to the order parameter variables of the type $\delta$.
The correlation functions (\ref{Correl-funct}) are obtained as
\begin{equation} \label{Corr-sources}
\chi _{l\bf q}^{\delta}(\theta_1,\theta_2;|\tau _1-\tau_2|)=
-\left[ \frac{
\delta ^2 \ln Z}
{\delta h^{\delta}_{\bf q}(\theta_1;\tau_1)\delta \bar{h}^{\delta}_{\bf q}
(\theta_2;\tau_2)}
\right] _{h,\Psi_{<},\bar{\Psi}_{<}=0}  \; .
\end{equation}
Putting slow modes to zero means symbolically that we want the response only 
from the fast modes, as defined in equation  (\ref{Correl-funct}).
 We consider only the static and long--wavelength limit. For that reason 
we will simply write $\chi(\theta_1,\theta_2)$  instead of $\chi_{\bf
q=0}(\theta_1,\theta_2; i\omega=0) $ and $h(\theta)$ instead of 
$h_{\bf q=0}(\theta; i\omega=0)$. The correlations with the nonzero ${\bf
q}$  and $\omega$ are related to  the dynamics of the 
collective modes, a problem which we do not study in this work.

\begin{figure}
\centerline{\epsfig{file=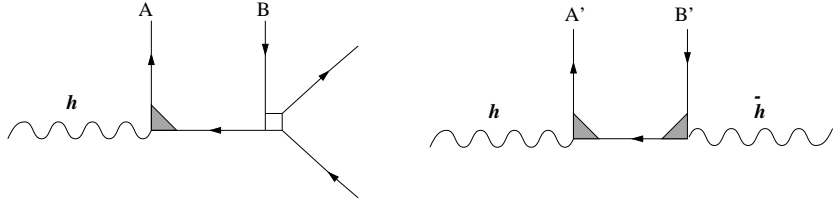,width=11cm}}
\caption{Tree diagrams conataining the source fields for AF.}
\label{Fig11}
\end{figure}

We now apply the Kadanoff--Wilson-Polchinski formalism to the action containing
terms (\ref{ext-fields}). 
 The procedure of collecting differential cumulants is analogous to 
what we  explained in  previous section, but now we treat $S\{ h\}$ 
terms together with the interaction part $S_I$. To obtain the 
correlation functions for 
$h\rightarrow 0$, it is sufficient to follow the renormalization of the
first two terms in powers of $h$ in the $h$--dependent part of 
 the effective 
action. They read:
\begin{eqnarray} \nonumber
\lefteqn{S\{ h\} _l=\oint d\theta_1 \oint d\theta_2\left[
\int _0^{\Lambda (l)} d\epsilon {\cal J} (\epsilon,\theta_1) \hat{\Delta}_{\bf
q}^{\delta}(\epsilon,\theta _1;\tau)
\right] z_l^{\delta}(\theta_1,\theta_2)\bar{h}^{\delta}(\theta_2)+
\mbox{h.c.}+} & \\
\label{Sh}
& + & \oint d\theta_1 \oint d\theta_2 \; \bar{h}^{\delta}(\theta_1)\chi
_l^{\delta}(\theta_1,\theta_2) {h}^{\delta}(\theta_2)
+\mbox{tree terms}\{ h\bar{h}\} \; .
\end{eqnarray}
The term with $\chi _l^{\delta}(\theta_1,\theta_2)$ contains 
no electronic variable: it results from  the elimination of all 
outer--shell electrons.
From the definition (\ref{Corr-sources}) one can see that 
 $\chi
_l^{\delta}(\theta_1,\theta_2)$  is just the 
susceptibility of type $\delta$. 
The  ``tree terms'' are the terms containing one outer--shell contraction,
two slow--electron fields and fields $h$ and $\bar{h}$. They are 
illustrated by  figure \ref{Fig11} for the AF channel.
The square symbolizes the effective interaction for antiferromagnetism 
$V_l^{AF}$.
We obtain it from the spin--spin interaction $U_{\sigma}$ (see appendix 
\ref{inv_interactions}) putting the particles 1 and 3 on the opposite sides 
of the square Fermi surface so that ${\bf k_1}-{\bf k_3}=(\pm \pi,\pm \pi)$
(as on  figure \ref{Fig10}):
\begin{equation} \label{VSDW}
V^{AF}_l(\theta_1,\theta_2)=-(XU)(\theta_1,\theta_2,\tilde{\theta}_1)\; .
\end{equation}
$\tilde{\theta}$ is a function of $\theta$ such that 
\begin{equation} \label{tilde}
{\bf
k}(\theta)-{\bf k}(\tilde{\theta})=(\pi,\pi)\; , 
\end{equation}
${\bf k}$ being on the 
square Fermi surface.
The tree terms for SC channel are analogous, but with different 
orientations of the arrows: in the vertex $z_l^{SC}$ both arrows point
 outwards and in the vertex $\bar{z}\l^{SC}$ both arrows point inwards.
The corresponding interaction is the familiar effective Cooper amplitude $V$
(\ref{V-marginal1}).
All ``tree terms'' in  equation (\ref{Sh}) are produced by the 
tree term of the Polchinski equation 
applied to the action with the $S\{ h\} _l$ terms. 

The coefficient
$z_l^{\delta}(\theta_1,\theta_2)$  is the effective vertex
of type $\delta$. The equation (\ref{ext-fields}) gives  
the initial conditions for $z$
\begin{equation} \label{z-incond}
z_{l=0}^{\delta}(\theta_1,\theta_2)=\delta _D(\theta_1-\theta_2),
\end{equation}
where $\delta _D$ is the Dirac function and for $\chi$
\begin{equation} \label{chi-incond}
\chi_{l=0}^{\delta}(\theta_1,\theta_2)=0.
\end{equation}

\begin{figure}
\centerline{\epsfig{file=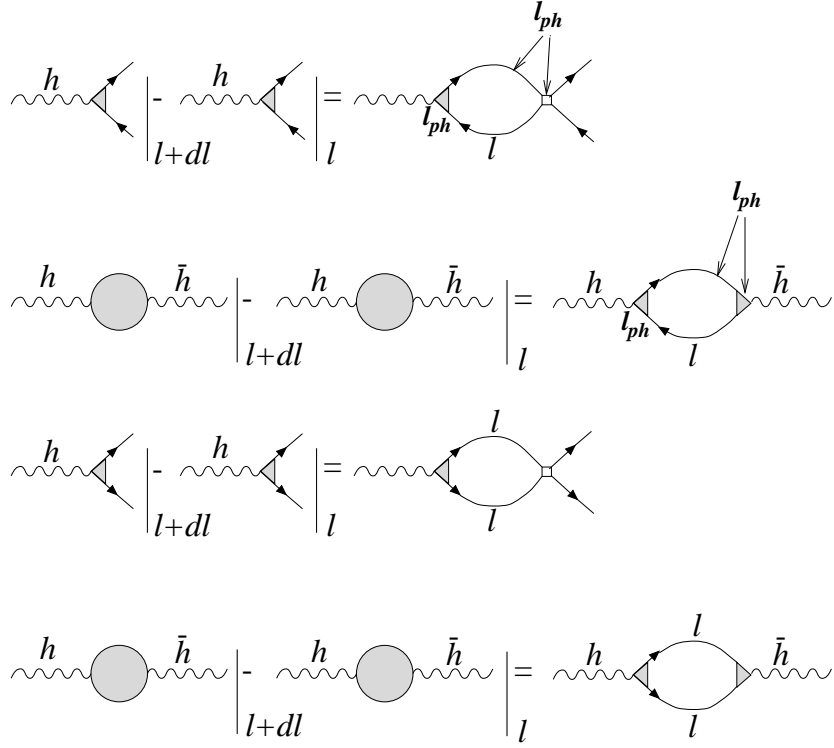,width=11cm}}
\caption{The recursion relations for the vertices and for the correlation 
functions of the superconducting and of the antiferromagnetic type.}
\label{Fig12}
\end{figure}

The differential flow of the triangular vertices $z_l$ and of the  
correlation functions $\chi$ is obtained from the loop diagram of 
the Polchinski equation applied to the tree terms in the action $S\{ h\} _l$.
For the AF channel the, cumulant with on-shell integration of the electrons 
A and B on  figure \ref{Fig11} gives the contributions to the vertex
$z_l^{AF}$ and the cumulant with electrons A' and B' on the shell contributes
to the susceptibility $\chi_l^{AF}$. A similar construction yields the 
renormalization of the vertex and of the susceptibility for the
superconductivity.
The resulting diagrams for the differential recursion relations for both 
channels are shown on figure \ref{Fig12}.
The corresponding flow equations write
\begin{equation} \label{flow-z}
\dot{z}_l^{\delta}(\theta_1,\theta_2)=-\oint d\theta \; 
{z}_{l_{\delta}}^{\delta}(\theta_1,\theta ) D_l^{\delta}(\theta )
V_{l_{\delta}}^{\delta}(\theta
,\theta_2) 
\end{equation}
and 
\begin{equation} \label{flow-chi}
\dot{\chi}_l^{\delta}(\theta_1,\theta_2)=\oint d\theta\; 
{z}_{l_{\delta}}^{\delta}(\theta_1,\theta ) D_l^{\delta}(\theta ) 
z_{l_{\delta}}^{\delta}(\theta
,\theta_2) 
\end{equation}
The scales $l_{SC}$ and $l_{AF}$ symbolize the scales $l_{pp}$ and $l_{ph}$ 
given by expressions (\ref{lpp}) and (\ref{lph}), with the total momentum
${\bf q}_{pp}=0$ and with the momentum transfer 
${\bf q}_{ph}=(\pi , \pi)$ (antiferromagnetic wavevector):
$$
l_{\delta}= \left\{
\begin{array}{l}
          l_{pp}\arrowvert _{{\bf q}_{pp}=0}=0 \hspace{10mm} \mbox{for} 
\hspace{3mm} \delta=\mbox{SC} \\
        l_{ph}\arrowvert _{{\bf q}_{ph}=(\pi,\pi)}=\ln \frac{\Lambda_0}
{\Lambda _l+2|\mu |}   \hspace{10mm} \mbox{for}\hspace{3mm} \delta=\mbox{AF}
\end{array} \right.
$$
We see that the renormalization of the antiferromagnetic correlation 
function is non--local in $l$ if the filling is not exactly one--half.
The function $D_l^{\delta}(\theta)$ is
\begin{equation} \label{D-SC}
D_l^{SC}(\theta)=\frac{1}{2}\sum _{\nu=+,-}{\cal J}(\nu\Lambda(l),\theta)
\end{equation}
for the superconducting channel and
\begin{equation} \label{D-DW}
D_l^{AF}(\theta)=\frac{1}{2}\frac{{\cal J}(-\Lambda(l),\theta)
}{1+|\mu|/\Lambda(l)}\; ,
\end{equation}
for the antiferromagnetism where only the negative shell $(\nu=-1)$
contributes to the flow. One sees that $D_l^{AF}(\theta)$
decays exponentially with $l$ for $\Lambda \ll
|\mu|$: in  the BCS regime  the correlation function for antiferromagnetism 
saturates with increasing $l$.

From the equations (\ref{flow-z}) and (\ref{flow-chi}) we see that 
information about the {\it symmetry} of the correlations is
determined  from the symmetry of the effective interactions :  functions
$D_l^{\delta}(\theta)$ have a total lattice symmetry, but the interactions 
$V_{l}^{\delta}(\theta ,\theta_2)$ can belong to any of the 
representations of the  crystal symmetry group, in our case the 
$D_4$ point group.
The decomposition of the interaction in terms of all basis functions of all 
irreducible representations of the $D_4$ group is discussed in detail in our 
previous paper.\cite{ZS_prb}
The diagonalization of the correlation functions 
${\chi}_l^{\delta}(\theta_1,\theta_2)$ gives the final answer about which 
correlations are dominant in both AF and SC channels.
The strength of the 
dominant correlations is associated to the maximal eigenvalues and
the corresponding eigenvectors determine the symmetry and the form 
of the microscopic fluctuating field.

\subsection{Discretization of renormalization group equations}

The interaction $U_l$ that we want to renormalize is a function of three 
continuous angular variables $\theta _1$, $\theta _2$ and $\theta _3$.
The beta function given by the equations 
(\ref{Beta.fctn}-\ref{Betaeh}, \ref{Xi-theta}, \ref{Pi-theta}) is a 
complicated function bilinear in $U$, and it does not seem 
possible to find 
an analytic solution for the flow of the interaction.
We thus use numerical method. 
For that purpose we cut the Brillouin zone in $m_i$
angular ($\theta$) patches (see figure \ref{Fig7}) and we assume that
the interaction is a function only of the patch indices 
$(i_1,i_2,i_3)$ of the three angles
$\theta _1$, $\theta _2$ and $\theta _3$. After the discretization of the 
interaction function 
the differential loops $\Xi$ and $\Pi$ become also functions of 
three indices:
\begin{equation} \label{Xi_i}
\Xi \{ U,U\}(i_1,i_2,i_3) =\sum _{i=0}^{m_i}
B_{pp}(i_1,i_2,i;l)U_{l_{pp}}(i_1,i_2,i)U_{l_{pp}}(i_3,i_4,i), 
\end{equation}
\begin{equation} \label{Pi-i}
\Pi \{ U_1,U_2\}(i_1,i_2,i_3)=\sum _{i=0}^{m_i} B_{ph}
(i_1,i_3,i;l)U_{1,l_{ph}}(i_1,i,i_3)
U_{2,l_{ph}}(i_4,i,i_2)
\end{equation}
with
\begin{equation} \label{B_ee}
B_{pp}(i_1,i_2,i;l)=\frac{-2}{(2\pi)^2}\sum _{\nu =+,-}
\int _{[i]} d\theta {\cal J}(\nu\Lambda,\theta)
\frac{
\Theta
\left(\nu\xi _{{\bf k}_{\nu}-{\bf q}_{pp}}
 \right) 
\Theta
\left(|\xi _{{\bf k}_{\nu}-{\bf q}_{pp}}|-\Lambda \right) 
}{1+\frac{\nu}{\Lambda}\xi _{{\bf k}_{\nu}-{\bf q}_{pp}}}
\end{equation}
and
\begin{equation} \label{B_eh}
B_{ph}(i_1,i_3,i;l)=\frac{2}{(2\pi)^2}\sum _{\nu =+,-}
\int _{[i]}  d\theta {\cal J}({\nu}\Lambda,  \theta,)
\frac{
\Theta
\left(-\nu\xi _{{\bf k}_{\nu}+{\bf q}_{ph}}
 \right) 
\Theta
\left(|\xi _{{\bf k}_{\nu}+{\bf q}_{ph}}|-\Lambda \right) 
}{1-\frac{\nu}{\Lambda}\xi _{{\bf k}_{\nu}+{\bf q}_{ph}}}.
\end{equation}
The total momentum and the momentum transfer become  discrete variables:
$${\bf q}_{pp}={\bf k}(i_1)+{\bf k}(i_2),$$
$${\bf q}_{ph}={\bf k}(i_1)-{\bf k}(i_3).$$
The integral $\int _{[i]}$ is over  $i$-th angular sector.

For a given number $m_i$ of patches the number of  {\it coupling constants}
is equal to the number of configurations of three indices for all 
four particles lying on the square Fermi surface. That is a very large number.
However because of the symmetry many of the coupling constants are identical.
The available symmetries are: (i) The symmetries of the $D_4$ point group 
(mirror, $\pi/4$--rotations); (ii) Time inversion symmetry $\cal T$, 
exchanging particles with  holes and vice versa 
(see appendix \ref{inv_interactions});
(iii) The exchange symmetry: 
it is allowed to exchange simultaneously (1,2) and 
(3,4) particles; (iv) The freedom of choice of the points at the edges 
of the Brillouin zone.
\begin{figure}
\centerline{\epsfig{file=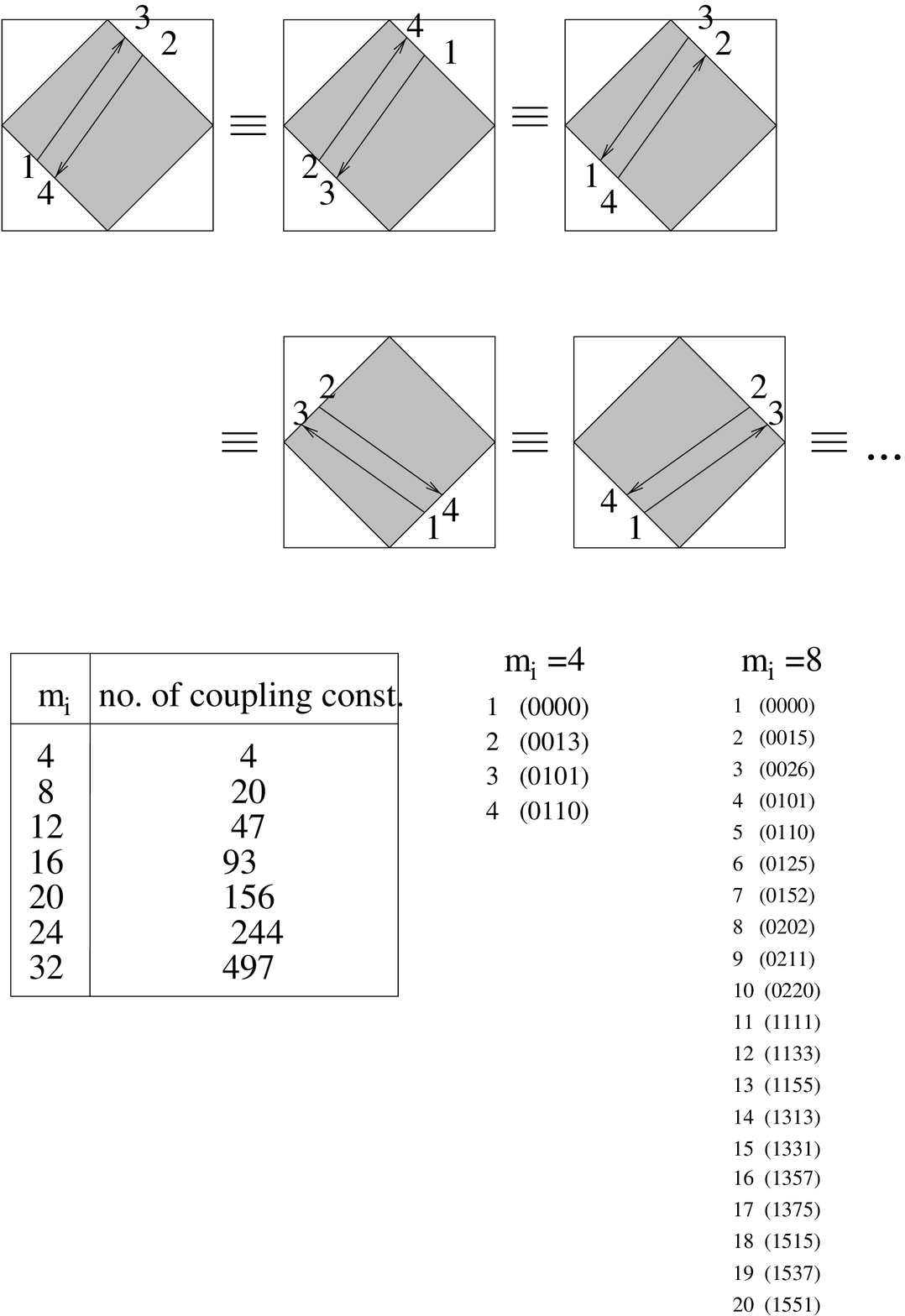,width=11cm}}
\caption{The figure shows how we reduce the number of  coupling constants
by applying  symmetry transformations. The dependence of the number of
independent coupling constants on the number of angular patches $m_i$ and 
the list of coupling constants for $m_i=4$ and $m_i=8$ are also shown.}
\label{Fig13}
\end{figure}
Figure \ref{Fig13} illustrates some of the symmetry operations  
 applied to one of the coupling constants. 
The same figure shows the relation between the 
number of patches and the corresponding number of different marginal 
coupling constants, and 
the list of the coupling constants for $m_i=4$ and $m_i=8$ .

The renormalization of the interaction as a function of three angles is now
represented by a set of coupled 
differential equations, one for each coupling constant.
In the same way we discretize the correlation functions 
${\chi}_l^{\delta}(\theta _1,\theta _2)$ and the vertices 
${z}_l^{\delta}(\theta _1,\theta _2)$. The equations 
(\ref{flow-z}) and (\ref{flow-chi}) become
\begin{equation} \label{flow-z-i}
\dot{z}_l^{\delta}(i_1,i_2)=-\sum _{i}
{z}_{l_{\delta}}^{\delta}(i_1,i) \bar{D}_l^{\delta}(i)
V_{l_{\delta}}^{\delta}(i
,i_2) 
\end{equation}
and
\begin{equation} \label{flow-chi-i}
\dot{\chi}_l^{\delta}(i_1,i_2)=-\sum _{i}
{z}_{l_{\delta}}^{\delta}(i_1,i) \bar{D}_l^{\delta}(i) 
z_{l_{\delta}}^{\delta}(i
,i_2)  
\end{equation}
with
\begin{equation} \label{D-i}
\bar{D}_l^{\delta}(i)\equiv \int _{[i]} d\theta D_l^{\delta}(\theta ).
\end{equation}
The initial conditions are the same as in the continuous case,
provided we replaced   the 
delta function by the Kronecker symbol divided by $m_i$:
$$\delta _D(\theta -\theta') \rightarrow \delta
_{i,i'}/m_i.$$

\subsection{Results and discussion}

We have integrated numerically the renormalization equations for all
coupling constants and for  
correlation functions and we have analyzed how the
 results change as functions of the initial interaction $U_0$ and of the 
chemical potential $\mu$.

\begin{figure}
\centerline{\epsfig{file=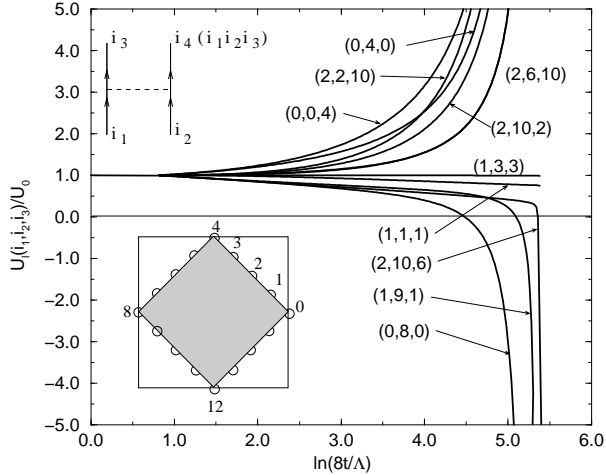,width=9cm}}
\caption{
The flow of a few typical (among 93) scattering amplitudes 
for a Fermi surface covered by 16 patches, for chemical potential 
$|\mu |=8t\exp (-7.8)$ and initial interaction $U=4t/3$.}
\label{Fig14}
\end{figure}

We first  look at  the renormalization flow of the
coupling constants. Figure \ref{Fig14} shows the flow of  several 
(among $93$) coupling constants for a  $M_i=$16--patches discretization;
the choice of the input parameters is $U_0=4t/3$ and
 $l_{\mu}\equiv \ln 8t/|\mu|=7.8$. The divergence happens at the critical
scale $\l_c\approx 5.3$. Approaching  this point, some of the coupling 
constants
increase and diverge, while some decrease and, after changing their sign, 
diverge to $-\infty$. Some do not change significantly upon renormalization 
and do not diverge. 
For example the coupling constant $U(0,m_i/2,0)$ diverges very strongly to 
$-\infty$. It is a typical interaction with singular Cooper channel 
$({\bf q_{pp}}=0)$ without nesting. Indeed, all coupling constants obeying 
 only Cooper condition  $({\bf q}_{pp}=0)$ and without logarithmic 
flow in p-h channel diverge 
to  $-\infty$. This is what we expected  since the p-p channel 
``pushes'' interactions downwards in a repulsive model. However, instead of
just decaying to zero, they continue to decrease towards $-\infty$ because
the Cooper 
amplitude obtains  {\em attractive} components  from p-h diagrams 
in, for example, the D--wave channel.
The coupling constants with  nesting between particles 1 and 3 or 1 and 
4 diverge to $+\infty$. Among the interactions with  nesting there are also
umklapp processes like (0,0,4) or (2,2,10) in  figure \ref{Fig14}.
The processes without divergence are those without any logarithmic instability
neither from the nesting nor from the Cooper logarithm.

%exemples of nested and Cooper amplitudes, coupling etc...

\begin{figure}
\centerline{\epsfig{file=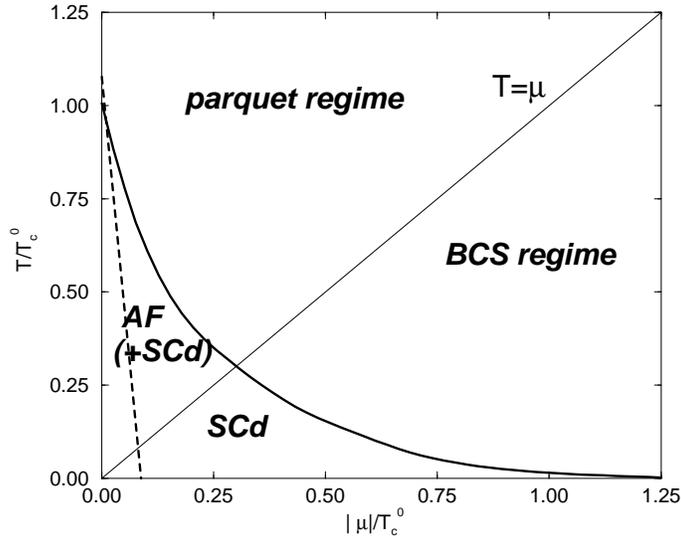,width=9cm}}
\caption{
The phase diagram. The solid line is the critical temperature $T_c^{RG}$
and the  dashed line is the temperature $T_c^{MF}$ .}
\label{Fig15}
\end{figure}
The critical scale $l_c$ depends on the  initial interaction and on the 
chemical potential. We associate the cutoff $\Lambda =\Lambda_0 \exp (-l_c)$
to the critical temperature $T_c^{RG}$. Figure \ref{Fig15} shows 
$T_c^{RG}$ as a 
function of the  chemical potential calculated for $m_i=32$ patches 
(497  coupling constants).
$T_c^{RG}$ decreases rapidly but never really  falls 
to zero : it becomes exponentially small far from  half 
filling, the regime analyzed  in ref.\onlinecite{ZS_prb}.
Our numerical calculations show that this form is universal if one measures 
$\mu$ in units of the critical temperature at  half filling $T_c^0$:
\begin{equation} \label{universTc}
T_c^{RG}=T_c^0 \times f\left( \frac{|\mu|}{T_c^0}\right) 
\end{equation}
where $f$ is the universal function visible on  figure \ref{Fig15}.
Thus $T_c$ depends on the interaction only through $T_c^0\equiv
8t\exp (-l_c^0)$, where
\begin{equation} \label{lcO}
(l_c^0)^2=C \frac{4t}{U_0}
\end{equation}
$C$ being a numerical constant, $C\approx 8.8$.
The dashed line in  figure \ref{Fig15} represents the critical 
temperature $T_c^{MF}$ that one obtains when taking into account 
only the last term of equation (\ref{Beta.fctn}):  $X\beta\{ XU,XU\} $.
This is the ``renormalization group'' version of the RPA summation, 
equivalent to the mean field for the  antiferromagnetism. We see now 
the main difference between the critical temperature in
 the mean field approximation and the result obtained with the
renormalization group: in the case of  weak doping, 
because of the {\it destructive} interference between  channels p-p and 
p-h, $T_c^{RG}$
is slightly reduced with respec to $T_c^{MF}$.
The ratio between the critical scales $l_c^{RG}$ and $l_c^{MF}$ (associated to 
RG and MF critical temperatures) do not depend on the interaction. 
Its value at half filling is
$
{l_c^{MF}}/{l_c^{RG}}=0.985 ,
$
which is not far from the value 0.981, calculated by Dzyaloshinskii and
Yakovenko using parquet equations.\cite{DzYak}
 $T_c^{MF}$ disappears completely at some threshold doping.
This means that the  physical mechanisms which reduce $T_c^{RG}$ near 
half filling, at  higher doping enhance $T_c^{RG}$   
keeping it always non--zero.

The straight line $T=\mu$ is roughly the crossover between the parquet and 
the BCS regimes.
If the instability occurs in the parquet regime, both p-p and p-h correlations 
are strongly enhanced near $T_c^{RG}$. On the other hand, in the BCS
regime only the p-p correlations are  critical.
To know wich which fluctuations are the 
most important at the instability, we need the renormalization of 
the correlation functions. However,  there is  a formal problem related 
to the 
fact that we are
performing the renormalization at $T=0$ and associating the 
cutoff to the temperature: the renormalization equations for the
antiferromagnetic vertex (\ref{flow-z-i}) and  the correlation function
(\ref{flow-chi-i}) ($\delta=AF$) are {\it retarded} in
$l$ of a quantity  $l-\ln \frac{\Lambda_0}{\Lambda _l+2|\mu |}$. 
If the interaction
$V_{l}^{AF}$ diverges at $l=l_c$, the divergence of the 
function $\chi_l^{AF}$ will be retarded. Since we can not go further
than $l=l_c$ in the renormalization  this divergence can not be seen in the 
present formalism. The cure is to work at finite temperature.
In that case one performs the full renormalization, up to $l=\infty$ and the 
final {\it fixed point} correlation functions are the ones at the given 
temperature. This is the procedure that we use in the following chapter.

However, in the zero temperature formalism on can still get some 
idea of what happens with different correlations  at $l=l_c$: The finite 
cutoff divergence of the
correlation functions for SC and AF are determined exclusively 
by the divergence of the effective interactions $V_{l}^{SC}$ and 
$V_{l}^{AF}$. 
Furthermore the symmetry of the correlation functions is also brought only 
by the symmetry of the effective interactions. It is thus reasonable to
assume that the dominant eigenvalue of the correlation function is 
driven mostly by the dominant attractive (negative)
eigenvalue of the corresponding effective 
interaction.
From the renormalization of the interaction (the set of coupling constants) 
we can deduce the flow of the effective interactions $V_{l}^{SC}$
and $V_{l}^{AF}$ as given by  equations (\ref{V-marginal1}) and
 (\ref{VSDW}). The diagonalization is straightforward because both interactions
are matrices whose rows and columns are labeled
by disretized angular variables.
Let's call the most attractive eigenvalues of $V_{l}^{SC}$ and $V_{l}^{AF}$
respectively  $V_{c}^{SC}$ and $V_{c}^{AF}$.
\begin{figure}
\centerline{\epsfig{file=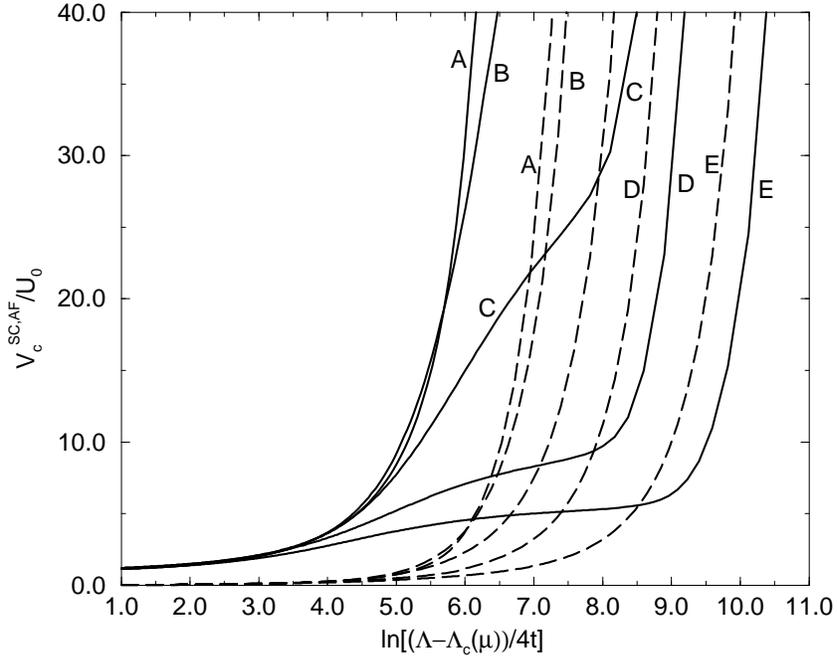,width=11cm}}
\caption{The flow of $V_{c}^{SC}$ (dashed line) and of 
$V_{c}^{AF}$ (solid line)
for $|\mu|/(4t)=$0(A); 0.00067(B); 0.0018(C); 0.0049(D); 0.0081(E).}
\label{Fig16}
\end{figure}
Figure \ref{Fig16} shows the flow of $V_{c}^{SC}$ and of 
$V_{c}^{AF}$ near the critical point
as a function 
of $\ln [(\Lambda-T_c^{RG}(\mu))/4t]$ for several values of  the chemical 
potential.
The critical temperature $T_c^{RG}(\mu)$ is adjusted for every value of $\mu$.
Solid lines represent the antiferromagnetic interaction $V_{c}^{AF}$.
The corresponding eigenvector  belongs to $A_1$. It is a standard S--wave.
The dashed lines represent the flow of  $V_{c}^{SC}$. 
Its eigenvector belongs to the $B_1$
representation ($D_{x^2-y^2}$--wave).
Both coupling constants are always enhanced by the renormalization, which
means that the correlation functions are always enhanced respectively to 
their value at $U_0=0$.
The possibility of the charge density wave instability is excluded: 
we have checked that all eigenvalues of the 
charge interactions $U_c$ (see appendix \ref{inv_interactions}) at $2k_F$
decay upon renormalization.
The competition between
 the divergences of $V_{c}^{SC}$ and $V_{c}^{AF}$ is clearly
visible in the figure. At  half filling the coupling $V_{c}^{AF}$ 
diverges faster than $V_{c}^{SC}$. As the chemical potential increases, 
both divergences are weaker but in  $V_{c}^{AF}$ an inversion of the 
slope is visible. 
This is the signature of the crossover from 
the parquet to the BCS regime. At half filling this crossover does not exist 
and the slope of $V_{c}^{AF}$ is always upwards. 
The lines labeled by (C) in the figure correspond to the critical 
temperature in the parquet regime. However, $V_{c}^{AF}$ starts to ``feel''
the proximity of the crossover: the critical scale is $l_c=6.025$ and the 
crossover occurs at about $l_x\sim l_{\mu}=7$. The divergence of 
$V_{c}^{AF}$ can still entail the divergence of the antiferromagnetic 
correlation function because the nesting is still relevant.
The Cooper amplitude $V_{c}^{SC}$  always has an upward slope and diverges at 
$T_c^{RG}$ because the p-p channel has a logarithmic instability for any 
doping.
Lines (D) and (E) are examples the flow in the BCS regime. 
After some saturation tendencies, $V_{c}^{AF}$ still diverges at $T_c^{RG}$. 
This divergence is only due to the p-p loop:
for a  choice of angles $\theta_1$ and $\theta_2$ such that
$\theta_2=\theta_1+\pi$, the function 
$V^{AF}(\theta_1,\theta_2)=V^{SC}(\theta_1,\tilde{\theta}_1)$, 
that diverge. The relation between $\theta_1$ and $\tilde{\theta}_1$ 
is given by equation (\ref{tilde}).
The interaction $V_{c}^{AF}$ is thus driven upwards by the Cooper channel.
It has no effect on the correlation function for the antiferromagnetism 
because its flow has disappeared together with the nesting.
This will become visible in the next section where 
we calculate the  temperature dependence of 
both correlation functions near the critical temperature.

\section{Finite temperature RG}

In the zero temperature formalism the 
flow of different quantities was of  physical interest.
In the finite temperature renormalization group, we are interested in the 
fixed point value of the correlation functions.
The temperature is taken as input parameter.
If $T$ is  larger than the critical cutoff $\Lambda _c$ 
(called $T_c^{RG}$ in the previous section),  the divergence
of the renormalization flow will disappear. 
Consequently we will be able to control the flow all the way 
down to the fixed point $\Lambda=0$. 
In the zero temperature formalism the effects of the 
elimination of the slow modes are neglected.
At  finite temperature all modes are integrated so that 
 contributions of the thermal electrons are also taken into account.
The other advantage of the finite temperature 
renormalization group is that we can explicitly follow the temperature 
dependence of the correlation functions for the 
superconductivity and the antiferromagnetism. 

Formally, the finite temperature procedure is the same as in the previous 
section
with the difference that the differential loops $\Xi$ and $\Pi$ have to be 
calculated at finite temperature. They now write
\begin{eqnarray} \nonumber
\lefteqn{\Xi \{ U,U\}(T,\theta _1,\theta _2,\theta _3)} & & \\
\nonumber 
&  = & \frac{-2}{(2\pi)^2}\sum _{\nu =+,-}
\int d\theta {\cal J}(\nu\Lambda,\theta)
\frac{
\left[ 1-f(\nu \Lambda)-f(\xi _{{\bf k}_{\nu}-{\bf q}_{pp}})
 \right] 
\Theta
\left(|\xi _{{\bf k}_{\nu}-{\bf q}_{pp}}|-\Lambda
 \right) 
}{\nu+\frac{1}{\Lambda}\xi _{{\bf k}_{\nu}-{\bf q}_{pp}}} \\
\label{Xi-theta-temp}
& & \times  U_{l_{pp}}(\theta_1,\theta_2,\theta)
U_{l_{pp}}(\theta_3,\theta_4,\theta)\; ,
\end{eqnarray}
\begin{eqnarray} \nonumber
\lefteqn{\Pi \{ U_1,U_2\}(T,\theta _1,\theta _2,\theta _3)} & & \\
\nonumber 
&  = & \frac{2}{(2\pi)^2}\sum _{\nu =+,-}
\int d\theta {\cal J}({\nu}\Lambda,\theta)
\frac{
\left[ f(\nu \Lambda)-f(\xi _{{\bf k}_{\nu}+{\bf q}_{ph}})
 \right] 
\Theta
\left(|\xi _{{\bf k}_{\nu}+{\bf q}_{ph}}|-\Lambda
 \right) }
{\nu-\frac{1}{\Lambda}\xi _{{\bf k}_{\nu}+{\bf q}_{ph}}} \\
\label{Pi-theta-temp}
& & \times  U_{1,l_{ph}}
(\theta_1,\theta,\theta_3)
U_{2,l_{ph}}(\theta_4,\theta,\theta_2)\; .
\end{eqnarray}
The function $f(\epsilon )$ is the Fermi distribution at  temperature $T$.
The finite temperature version of the flow equations for the vertices $z$ and 
for the correlation functions $\chi$ are again given by equations 
(\ref{flow-z}) and (\ref{flow-chi}) but with modified $D^{SC}$
\begin{equation} \label{D-SC-T}
D_l^{SC}(T,\theta)=\sum _{\nu=+,-}{\cal J}(\nu\Lambda,\theta)
\tanh \left( \frac{\nu \Lambda}{2T} \right)
\end{equation}
and $D^{AF}$
\begin{equation} \label{D-DW-T}
D_l^{AF}(T,\theta)=
\sum _{\nu=+,-}{\cal J}(\nu\Lambda,\theta)
\frac{ \left[ f(\nu \Lambda)-f(2|\mu| -\nu \Lambda) \right]
\Theta
\left(|2|\mu| -\nu \Lambda|-\Lambda
 \right)}
{2(\nu -\frac{|\mu |}{\Lambda})} \; .
\end{equation}
We see that now both shells $\nu=+,-$ contribute   to 
$D_l^{AF}$ for $\Lambda >|\mu|$, unlike in the zero temperature case where 
$\nu=+$ contributions were forbidden by the Fermi distribution.
In the discretized version of the flow equations, one calculates $\Xi$ 
and $\Pi$
from expressions (\ref{Xi_i}) and (\ref{Pi-i}), but with 
\begin{eqnarray} \nonumber 
\lefteqn{B_{pp}(i_1,i_2,i;l,T)} \\
& = & \frac{-2}{(2\pi)^2}\sum _{\nu =+,-}
\int _{[i]} d\theta {\cal J}(\nu\Lambda,\theta)
\frac{
\left[ 1-f(\nu \Lambda)-f(\xi _{{\bf k}_{\nu}-{\bf q}_{pp}})
 \right]
\Theta
\left(|\xi _{{\bf k}_{\nu}-{\bf q}_{pp}}|-\Lambda \right) 
}{\nu+\frac{1}{\Lambda}\xi _{{\bf k}_{\nu}-{\bf q}_{pp}}}
\label{B_ee_T}
\end{eqnarray}
and
\begin{eqnarray} \nonumber 
\lefteqn{B_{ph}(i_1,i_3,i;l,T)} \\
& = & \frac{2}{(2\pi)^2}\sum _{\nu =+,-}
\int _{[i]}  d\theta {\cal J}({\nu}\Lambda, \theta)
\frac{
 \left[ f(\nu \Lambda)-f(\xi _{{\bf k}_{\nu}+{\bf q}_{ph}})
 \right]
\Theta
\left(|\xi _{{\bf k}_{\nu}+{\bf q}_{ph}}|-\Lambda \right) 
}{\nu-\frac{1}{\Lambda}\xi _{{\bf k}_{\nu}+{\bf q}_{ph}}}\; .
\label{B_eh_T}
\end{eqnarray}
The equations for finite temperature $z_l^{\delta}(T,i_1,i_2)$ and 
$\chi_l^{\delta}(T,i_1,i_2)$ are (\ref{flow-z-i}) and (\ref{flow-chi-i}) with
$D_l^{\delta}(T,i)$ calculated from (\ref{D-i}), but using (\ref{D-SC-T})
and (\ref{D-DW-T}).

To find each point of the phase diagram, we have to find the fixed--point
$(\Lambda \rightarrow 0)$ value of the maximal eigenvalues 
$\chi _c^{\delta}(T,l)$ of the correlation functions 
$\chi _l^{\delta}(i_1,i_2)$; $(\delta =SC,AF)$. 
It means that the 
complete renormalization from $l=0$ to $l\rightarrow \infty$ has to be 
done for each temperature.
\begin{figure}
\centerline{\epsfig{file=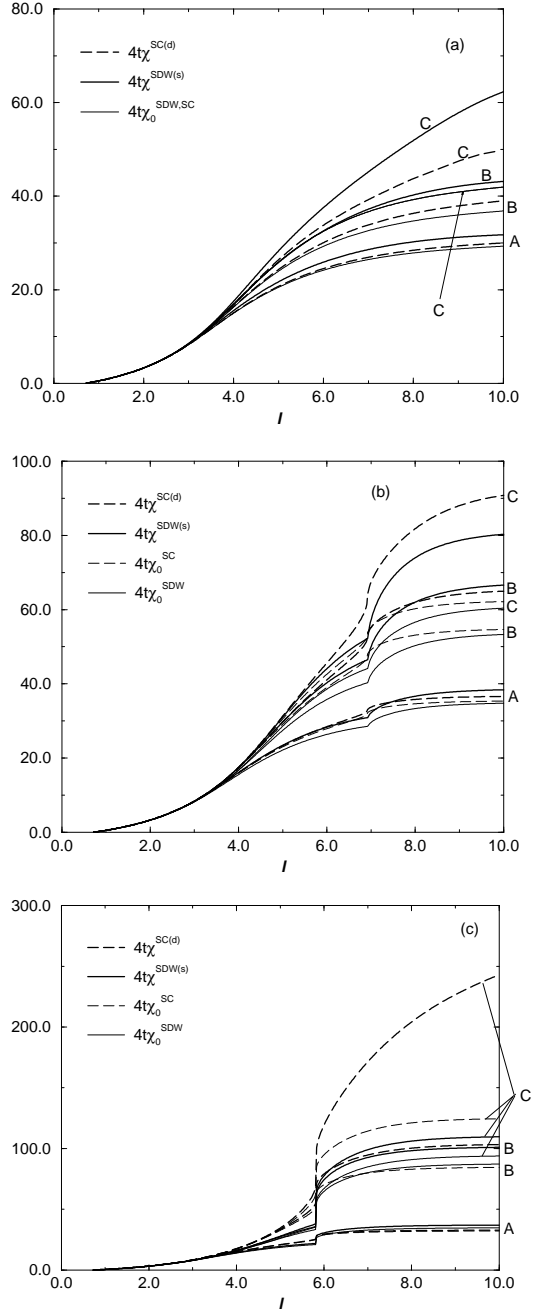,width=7cm}}
\caption{The flow of the correlation functions
with  interaction (thick lines) and without interaction (thin lines)
(a): at  half filling for $T/4t=0.03$(A), 0.0204(B), 0.0163(C); (b): 
at $|\mu|/4t=0.002$ for  $T/4t=0.0228$(A), 0.0108(B), 0.0086(C); and 
(c): at 
$|\mu|/4t=0.006$ for $T/4t=0.03$(A), 0.006(B), 0.0026(C).}
\label{Fig17}
\end{figure}
The flow of the quantities
 $\chi _c^{SC}(T,l)$ and $\chi _c^{AF}(T,l)$ is 
shown in  figure \ref{Fig17} for several temperatures and three different 
values of the chemical potential. The susceptibilities for the non--interacting
$(U=0)$ case are also shown. For all calculations the initial interaction was 
$U_0=4t/3$ and we have cut the Brillouin zone into $m_i=32$ patches.
The symmetry of the dominant superconducting correlations is for all cases
$B_1$ (which transforms as $d_{x^-y^2}$) and the dominant 
antiferromagnetic correlations have $A_1$ (s) symmetry. They correspond to
the symmetries of the strongest attractive components of the 
effective interactions $V^{SC,AF}$ found in the previous section.

\begin{figure}
\centerline{\epsfig{file=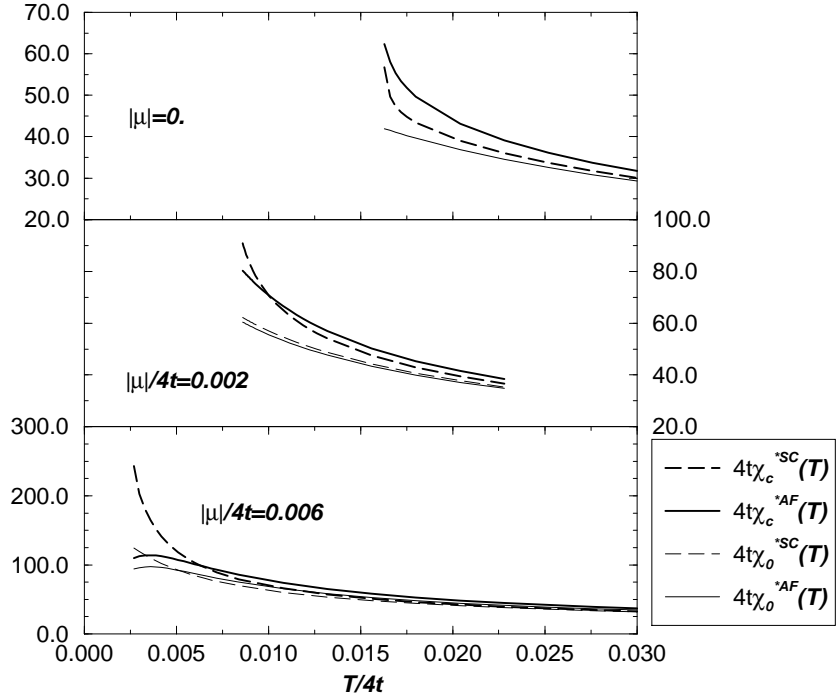,width=11cm}}
\caption{The temperature dependence of the fixed--point correlation functions
for three different values of the chemical potential.}
\label{Fig18}
\end{figure}

Let's concentrate first on figure \ref{Fig17}(a) that shows the flow at
 half filling. 
The entire flow is in the parquet regime: the nesting is perfect.
In the beginning of the flow where $\Lambda _l\gg T$ all correlation 
functions, bare or with correlations,
scale as if the temperature was zero, i.e. like 
$\ln ^2(\Lambda _0/\Lambda) =l^2$. As the cutoff approaches  the temperature,
the flow starts to saturate. At the same time the effects of the interaction 
become more and more visible as we decrease the temperature.
For all temperatures $\chi _c^{AF}$ and $\chi _c^{SC}$ are enhanced from their
bare values $\chi _0^{AF}$ and $\chi _0^{SC}$ 
that are equal at  half filling. 
As we approach the temperature $T\approx 0.0163$ from above, the difference
between the bare and the interacting cases  increases rapidly; 
we interpret this temperature as the critical temperature.
We have approximated the fixed point values $\chi _c^{*\delta}(T)$ and 
$\chi _0^{*\delta}(T)$
of $\chi _c^{\delta}(T,l)$ and $\chi _0^{\delta}(T,l)$
with their value for $l=10$ (the corresponding energy $\Lambda_l$ 
is   much smaller than any physical energy scale). Figure 
\ref{Fig18}(a) shows the temperature dependence of the fixed point values
at half filling. The bare susceptibility scales  as
$\ln ^2(\Lambda _0/T)$. The interaction makes both susceptibilities diverge, 
but  the antiferromagnetic one diverges first.

Now we increase the chemical potential to $|\mu|/4t=0.002$ (figure 
\ref{Fig17}(b)). The beginning of the flow, where $\Lambda _l\gg |\mu|$,
is still square--logarithmic. When $l$ becomes close to $\l_{\mu}=6.9$ 
the bare antiferromagnetic correlations  start to be weaker because we 
approach the crossover from the parquet to the BCS regime.
The non--analytic 
point is at $l=l_{\mu}$; at this point the flow equations (\ref{flow-z-i}) and
(\ref{flow-chi-i}) have a peak because of the van Hove singularity in 
$D_l^{\delta}(\theta )$ at $\theta=0,\pi/2,\pi,3\pi/2$.
Again, as we approach the critical temperature the effects of the interaction
become stronger and stronger so that  the difference between 
$\chi _c^{\delta}(T,l)$ and $\chi _0^{\delta}(T,l)$ increases more 
and more. The temperature dependence of the fixed point values for  all 
correlation functions for the present case are shown in figure 
\ref{Fig18}(b).
The critical temperature is $T_c^{RG}/4t=0.0075$,which is higher than 
$|\mu|/4t=0.002$. This means that the instability is still in the parquet 
regime, but not too deep: the proximity of the crossover already affects the 
antiferromagnetic correlations which start to loose their strength with respect 
to the superconducting correlations near the instability. However, both are
still strongly enhanced  and their flow is 
 dominated by the parquet
part $(l<l_{\mu})$ for all temperatures $T > T_c^{RG}$. 

Let's increase further the chemical potential to $|\mu|/4t=0.006$ (figure 
\ref{Fig17}(c)). The  flow of the antiferromagnetic correlations 
saturates in the BCS regime $(l>l_{\mu}=5.8)$, but both correlation functions 
SC and AF remain enhanced from their bare values. In the temperature 
dependence of their fixed point values (figure \ref{Fig18}(c)) one sees that 
only the superconducting instability is possible. The critical 
temperature is lower than the chemical potential, i.e. the instability is in 
the BCS regime,
 in which the p--h part of the flow is negligible. The antiferromagnetic
susceptibility even starts  to decrease with the temperature when 
$T \lessim \mu$.
This happens because we do not adjust the wave vector of the SDW to the best 
nesting (incommensurate
SDW) but we keep it for simplicity at $(\pi,\pi)$. Nothing drastic would 
happen even if we have taken small deviations of the best nesting 
wave vector from $(\pi,\pi)$: the susceptibility would saturate as
 the temperature decreases because the differential p-h bubble 
decays in the 
BCS regime {\it at any wave vector} with positive power of $\Lambda $.

The universal function $f(|\mu|/T_c^0)$ (equation \ref{universTc}),
which determines the dependence of 
the critical temperature on the chemical potential, is practically the same as 
the one obtained in the zero temperature formalism.
The final phase diagram is the one on figure \ref{Fig15}.
At  half filling the antiferromagnetic fluctuations are dominant over the 
superconducting ones but both correlation functions diverge at the 
critical temperature.
Upon doping
the antiferromagnetic correlations loose their strength while the 
superconducting correlations remain strongly divergent.
The divergence of AF correlation functions is completely suppressed if
the critical temperature is in the BCS regime.

\section{Conclusion}

We have formulated the exact Kadanoff--Wilson--Polchinski (KWP)
renormalization group for a general problem of  interacting
fermions on a two--dimensional  lattice. 
In principle the generalization to higher dimensions is trivial.
The procedure of  the KWP renormalization scheme is to integrate out
successively the degrees of freedom starting from high energies
and to follow 
 the renormalization of {\it all}
terms in the effective action.
We parametrize the renormalization by a high--energy cutoff 
$\Lambda\equiv \Lambda_0\exp(-l)$ 
determining the ring $\pm \Lambda$ around the Fermi surface.
In order to take the whole Brillouin zone into account,
the cutoff $\Lambda$ 
is taken to be equal to the bandwidth $(\Lambda_0=B.W.)$ 
at the beginning of the renormalization $(l=0)$ . 
As one proceeds with the mode elimination,
 vertices 
of all orders  are created.
To follow the exact renormalization of the effective action we need  to
know the flow (the dependence on $l$) of all vertices. 
The Polchinski equation (equations \ref{Polchinski1}, 
\ref{Polchinski2} and 
figure \ref{Fig3}) determines the differential flow of all vertices
as {\it functions} of energy--momenta.
In principle  the fixed point solution $(l\rightarrow \infty)$ of this 
equation gives us the exact connected Green functions of the model.
\cite{Polchinski84,Morris94} Clearly, the exact integration of  
Polchinski equation is impossible and for concrete calculations we have to
truncate the effective action.

The truncation at the sextic term (at the three--particle interaction term) 
generates
the one--loop renormalization group for the two--body
interaction. The truncated effective action is 
given by  expression (\ref{action1}). Its renormalization is determined 
by the flow equation for  the two--body interaction
and for the 
self--energy.
The flow equation for the interaction  $U_l$ is 
made of all one--loop diagrams  bilinear in $U_l$ as 
shown on  figure \ref{Fig4} and by  equation (\ref{Beta.fctn}).
Note that $U_l$ is renormalized as a function of three energy--momenta
(the fourth is conserved), i.e. this is a {\it functional} renormalization 
group. 
The $\beta$ function (\ref{Beta.fctn}) contains the contributions 
from the particle--particle (p-p)
$(\beta _{pp})$ and the particle--hole (p-h)  diagrams $(\beta _{ph})$. 
The first term 
is  called the BCS contribution in the literature, 
the next three terms are the zero sound (ZS)
contribution, and the last term is the ZS' contribution to
 the differential flow.
The flow equation for the interaction is not local in $l$ as one can see 
from  equations (\ref{Xi}) and (\ref{Pi}) for differential p-p and p-h 
bubbles: at some step $l$ of the renormalization, $U_l$ is 
renormalized by the values of $U$ at former steps
 $l_{pp}({\bf k,k_1,k_2})$ and 
$l_{ph}({\bf k,k_1,k_3})$. This non--locality is the price we have to pay if 
we want to keep all contributions, logarithmic or not, that renormalize
the interaction. 
In this way one takes correctly
into account, for example, the p-h flow because of the imperfectly 
nested Fermi surface.
The standard (local) Wilsonian 
RG \cite{Shankar} that takes into account only dominant logarithmic 
diagrams (those with $l=l_{pp}=l_{ph}$) can give useful results only for 
the perfectly nested (but not square) 
Fermi surfaces or the Fermi surfaces  far from being 
nested, so that the p-h part is negligible.

We have applied the one--loop renormalization group to the Hubbard model
on a square lattice near  half filling.
The interaction function $U$ that we renormalize is dependent on the angular 
($\theta$)
position of three momenta on the square Fermi surface 
(the fourth one is conserved). All radial momentum dependencies and energy 
($\omega$) dependencies are irrelevant to the Fermi liquid scaling.
It is important that we allow variables $\theta$  of the interacting 
particles to be anywhere on the square Fermi surface and not only in the
 configurations 
which give  perfect nesting or zero total momentum (see figure \ref{Fig6}).
This means that we do not limit ourselves  to the leading logarithmic 
parts of the flow but that we take  all non--logarithmic contributions 
into account. 

From the explicit scale dependence of the differential flow for $U$ we see 
two renormalization regimes (see figure \ref{Fig9}).
In the first regime, $\Lambda _l > |\mu|$. We call it  the parquet regime
because  both p-p and p-h contributions are important. The 
other regime exists in the non--half
filled case  when  $\Lambda _l < |\mu|$. There, only p-p loops have a strong
logarithmic flow while the p-h part decays to zero. We call this regime the 
BCS 
regime.
The effective phase space $(|\xi _0({\bf k})|<\Lambda )$ in the parquet 
regime is open so that the nesting is relevant (see figure \ref{Fig2}), 
while in the BCS regime the phase space is a closed regular ring of degrees 
of freedom around the Fermi surface so that  perfect nesting is impossible 
(see figure  \ref{Fig5}). The flow in the parquet regime is characterized by 
a strong coupling between the p-p and the p-h channels of renormalization. 
This 
coupling comes into play over the interactions that have a strong flow from
both p-p and p-h diagrams. For the case of the (nearly) square Fermi surface 
these are
all interactions between electrons from 
 opposite sheets of the Fermi surface.

The leading correlations in the Hubbard model are expected to be
antiferromagnetic and/or superconducting. To give a precise answer to
 this question, we use Kadanoff--Wilson--Polchinski procedure to construct the 
renormalization--group equations for the {\em angle resolved} 
correlation functions $\chi _l^{AF}(\theta _1, \theta_2)$ and 
$\chi _l^{SC}(\theta _1, \theta_2)$ for  antiferromagnetism 
and  superconductivity, defined
by  equation (\ref{Correl-funct}).
At a given step $\Lambda (l)$ of the renormalization these correlation 
functions 
measure the linear response of the electrons outside the shell $\pm \lambda$
around the Fermi surface. We take the static long-wave limit. The 
renormalization equation for $\chi _l^{AF, SC}$ is (\ref{flow-chi}).
The renormalization of the correlation functions depends on the renormalization
of the vertices $z^{AF,SC}$ (eq. \ref{flow-z}).
Furthermore, from  equations (\ref{flow-z}) and (\ref{flow-chi}) 
one sees that the flows of the susceptibilities and of the vertices 
depend on the flows of the corresponding {\em effective interactions}
$V_l^{SC}$ and $V_l^{AF}$, given by (\ref{V-marginal1}) and (\ref{VSDW}) 
respectively.

The flow equations for the interaction $U_l$ (\ref{Beta.fctn}), for the 
vertices $z_l^{SC,AF}$ (\ref{flow-z}) and for the correlation functions 
$\chi _l^{SC,AF}$ (\ref{flow-chi}) can be integrated numerically if we 
discretize their $\theta$--dependence. The coupling function is then 
approximated by a set of coupling constants. 
The vertices and correlation functions become discrete matrices.
Using physical and geometrical symmetries we reduce  number of the 
coupling constants to  a set of the independent ones 
(see figure \ref{Fig13}).
The functional renormalization--group equations become a set of equations, 
one for each coupling constant and for each matrix element of $z_l^{SC,AF}$
and $\chi _l^{SC,AF}$.

We have solved the renormalization equations for up to $m_i=32$ angular 
patches.
The typical flow of the coupling constants is shown on  figure 
\ref{Fig14}. There is a critical scale for which a large number of 
coupling constants diverge.
We associate the corresponding cutoff  to the critical temperature 
$T_c^{RG}$.
Its dependence on the chemical potential is shown on  figure \ref{Fig15}
(solid line) together with the RPA result (dashed line).
At the line $T_c^{RG}(\mu )$ the electronic correlations are strongly enhanced.
The type and the form of the corresponding microscopical fluctuating 
fields are given by the dominating eigenvalues (and their eigenvectors) 
of the correlation matrices 
$\chi _l^{SC,AF}$. These are 
 determined by the dominant attractive eigenvalues $V_c^{SC}$ and 
$V_c^{AF}$ of the effective interactions.
For all values of the chemical potential studied in this work, the 
eigenvalue $V_c^{SC}$ corresponds to $d_{x^2-y^2}$ (or $B_1$) singlet 
superconductivity while $V_c^{AF}$ is an  s--wave ($A_1$ representation).
The flow of the interactions $V_c^{SC}$ and $V_c^{AF}$  in the vicinity 
of the critical point $\Lambda=T_c^{RG}$ is shown on  figure \ref{Fig16}.
At  half filling $V_c^{AF}$ is dominant. Upon doping, the divergence of 
$V_c^{AF}$ loses its strength and the divergence of $V_c^{SC}$ becomes 
dominant.

To determine more precisely the dominant fluctuations near $T_c^{RG}$, we 
have done one further step in the renormalization--group formalism:
we have introduced the temperature explicitly into the flow equations.
In this formalism the cutoff $\Lambda$ has no more a physical meaning of the 
effective temperature.
At given temperature $T$ the physical information is contained in the 
fixed point $(\Lambda \rightarrow 0)$ of the correlation functions.
This extension of the formalism was necessary because in the 
zero--temperature procedure it was not possible to have the divergence 
of $\chi _l^{AF}$ at the same scale $l_c$ as the divergence of the 
coupling $V^{AF}$: for any nonzero chemical potential the flow of 
$\chi _l^{AF}$ has a finite retardation in $l$ (see equations 
(\ref{flow-z}) and (\ref{flow-chi}). Thus $\chi _l^{AF}$ diverges later,
at $l>l_c$.
In the finite temperature formalism $\chi _l^{AF}$, $\chi _l^{SC}$,
$V_l^{AF}$ and $V_l^{SC}$ all diverge  at the same temperature. The price 
to pay is that for each temperature we have to integrate the complete flow
all the way from $l=0$ to $l=\infty$ and to follow how the result changes 
with the temperature.

The flow of the dominant eigenvalues of $\chi _l^{AF}$ and $\chi _l^{SC}$
at a few different values of the temperature is shown on figure 
\ref{Fig17} for three different values of the chemical potential.
Both correlation functions are always enhanced with respect to their bare 
$(U=0)$ values.
The temperature dependence of the fixed point correlation functions is 
shown on figure \ref{Fig18}. The critical temperature $T_c^{RG}$ 
found by the finite temperature method is practically the same as the one 
found by the zero--temperature calculations, but now we are able to follow
 explicitly the enhancement of the correlations of both types in the 
vicinity of the instability.
It is clearly visible how the doping favors superconductivity and 
how the divergence of $\chi^{*AF}$ is completely suppressed if the 
instability is in the BCS regime, i.e. if $T_c^{RG}<|\mu|$.
This result justifies the phase diagram on figure \ref{Fig15}.

In the low--doping regime, both correlations are strongly enhanced and 
the low--temperature phase can be in principle  a mixture of both 
(quasi-)long--range--orders, with the superconducting component falling 
to zero at  half filling.
The instability is in the parquet regime: the critical fluctuations are a
 mixture of two fluctuating channels and can not be treated by an 
effective mean--field theory like BCS or RPA. In other words, the parquet 
regime is deeply non Migdalian so that the vertex corrections 
are as important as the p-p loops. The vertex corrections
can not be seen as small any more 
but have to be taken at all orders, just as the p-p diagrams, and together 
with other p-h loops.

The situation is less complicated in the BCS regime.
There, a low energy effective action can be constructed so that only the p-p
diagrams contribute to the instability while the p-h parts (antiferromagnetic
tendencies) are irrelevant. The attractive d-wave component of the
 Cooper amplitude in the LEEA is due to the higher energies 
($\epsilon >|\mu|$) where the p-h diagrams are important. 
In the BCS regime only superconductivity is possible as a 
low--temperature order.

As we
are considering a two--dimensional system, one should be careful about
the interpretation of $T_c$: in the case of magnetism, this indicates the
onset of well--defined finite--range correlations. For weak interactions,
this is typically a very well--defined crossover.\cite{schulz_89} In the
case of pairing $T_c^{RG}$ can be identified with the onset of 
quasi--long--range
order. Furthermore, the line between $AF(SCd)$ and $SC$ phases
on figure \ref{Fig15} is only partially determined in our calculations:
we only know  that at temperature near $T_c^{RG}$, this line is close to the 
crossover line $T=|\mu|$, but at lower temperatures we can not say anything
about its position.

It is difficult to discuss the experimental results from the point of view of 
our phase diagram. First of all, the one--loop renormalization--group is a weak
coupling perturbative method while the interactions in the copper--oxide 
superconductors are  moderate--to--strong.
For that reason our phase diagram can be compared to the 
experiments only qualitatively.
Furthermore, 
in our calculations we have neglected self-energy corrections, which are in
Polchinski's formalism given by Hartree--Fock--like terms with renormalized
$\omega$-- and {\bf q}--dependent vertices (eq. \ref{alpha.fctn}). 
The broadening and redistribution
of the spectral weight of the quasiparticles is then determined by the
dynamics of the vertex, which is irrelevant 
and is therefore neglected. One should however notice that at the
two--loop level self--energy effects  become important, as known from the
one--dimensional case.\cite{Rev1} In that sense, our $T_c^{RG}$ should be
understood as a temperature where the effects of interactions start to change
strongly not only the two--particles correlations, but the single particle
properties as well.
For that reason it seems natural to associate the temperature $T_c^{RG}$
to the crossover temperature $T_{co}$ found in the cuprates.
The parquet regime would then correspond roughly 
to the under-doped situation
and the BCS regime to the over-doped regime. 

The ``phase'' $AF(SCd)$ corresponds then to the antiferromagnetism and
to the pseudo-gap regime: the antiferromagnetic 
correlations 
and the localization tendencies are there accompanied more and more 
with the 
superconducting correlations as we approach the crossover line $T=|\mu|$.
We expect that the  critical temperature for antiferromagnetism and 
for superconductivity in this regime is lower than $T_c^{RG}$ 
because of 
the self--energy corrections. In other words, at 
temperature $T_c^{RG}$ in the parquet regime, 
the local antiferromagnetic moments
 and d--wave singlets are created with  finite 
correlation lengths. This gives rise to the pseudogap in both spin and 
charge responses, together with the precursors
of both antiferromagnetism and d--wave superconductivity.
The  
long--range--order between superconductivity and  antiferromagnetism 
 is perhaps  absent due to the fact that both types of fluctuations 
are strong. That is the 
central idea of the SO(5) models \cite{so5} for the high-$T_c$ 
superconductivity. 
In that language our $T_c^{RG}$
would play the role of the mean--field  critical temperature.

In the BCS regime only the superconducting fluctuations are critical.
We associate thus the phase $SCd$ to the overdoped regime.
From large N arguments\cite{Shankar,ZS_prb} we know that the self--energy
corrections  disappear  as $T_c/t$ if the Fermi surface is not nested.
This is the case in the BCS regime where the nesting processes are 
irrelevant.
Consequently,  the critical temperature in this regime is  well 
approximated by 
 $T_c^{RG}$. This is in agreement with the experiments: 
in the overdoped regime the crossover
temperature  $T_{co}$ is equal to the critical temperature for the 
superconductivity. 
Finally, mean-field
arguments\cite{these} suggest that one expects an incommensurate SDW 
(ICSDW) only in
the BCS regime and only where  the imperfect nesting is still strong, 
i.e. not far from the crossover $T=|\mu|$.
However, the precision
of our calculation (we cut the Brillouin zone into up to 32
$\theta$-patches) is not sufficient to check whether a magnetic correlation
function diverges  at some incommensurate wave vector.
In any case, the incommensurate 
SDW and d-superconductivity are not in competition 
because they appear at different places on the Fermi surface: SCd in the 
corners and ICSDW on the flat parts; one thus expects their coexistence.

Altogether, the phase diagram on figure \ref{Fig15} has important
similarities to the experimental phase diagrams. 
The one--loop renormalization--group,
taking into account  electron--electron and 
electron--hole processes on the same footing
reveals the essence of the physics of a doped  half filled
band  of correlated electrons.

\acknowledgements
Important comments of P. Nozi\`eres are acknowledged.
D. Z. thanks J. Schmalian for interesting discussions and K. H. Bennemann
for his hospitality at the 
Institut f\"ur Theoretische Physik der Freien Universit\"at Berlin.
Laboratoire de Physique Th\'eorique et Hautes Energies is Laboratoire 
associ\'e au CNRS UMR 7589. The work of D.Z. during his stay at
Institut f\"ur Theoretische Physik der Freien
Universit\"at Berlin was done in the framework of 
an Alexander von Huboldt fellowship.

\appendix
\section{Interaction $U(1,2,3)$ and its symmetries} 
\label{inv_interactions}
The most general spin--rotation invariant interaction term can be written in 
several ways. One way is in terms of charge--charge and spin--spin interactions
\begin{equation} \label{SpinCharge}
U_c(K_1,K_2,K_3)\bar{C}(K_2,K_4)C(K_3,K_1)+U_{\sigma}(K_1,K_2,K_3)
\bar{\bf S}(K_2,K_4)\cdot {\bf S}(K_3,K_1)\; ,
\end{equation}
where $C$ et $S_i$ are 
\begin{equation} \label{CSoper}
C(K_3,K_1)\equiv \sum _{\sigma} \bar{\Psi} _{\sigma K_3}\Psi _{\sigma
K_1} \hspace{10mm};\hspace{10mm} S_i(K_3,K_1)=\sum _{\sigma \sigma '}\bar{\Psi}
_{\sigma K_3} \sigma ^i_{\sigma \sigma '} \Psi _{\sigma '
K_1}\; .
\end{equation}
The summation over all three energy--momenta $(K_1,K_2,K_3)$ is assumed and
$K_4=K_1+K_2-K_3$.
On the other hand, the interaction can also be written as a sum of one term
with equal $(\sigma =\sigma ')$ and one with opposite $(\sigma =-\sigma ')$
spin quantum numbers, with corresponding coupling functions named
$U_{\parallel}(K_1,K_2,K_3)$ and $U_{\perp}(K_1,K_2,K_3)$:
\begin{equation}\label{par-anipar} 
U_{\parallel}(K_1,K_2,K_3)
\bar{\Psi}_{\sigma K_3}\bar{\Psi}_{\sigma K_4}{\Psi}_{\sigma K_2}
{\Psi}_{\sigma K_1} 
+U_{\perp}(K_1,K_2,K_3)
\bar{\Psi}_{\sigma K_3}\bar{\Psi}_{-\sigma K_4}{\Psi}_{-\sigma K_2}
{\Psi}_{\sigma K_1}
\end{equation}
with the summation over  spin indices assumed.
Spin--rotation invariance allows us to write the interaction part of the action
as a sum of the singlet $(|\vec{\sigma }+\vec{\sigma '}|=0)$ and triplet
$(|\vec{ \sigma }+\vec{\sigma '}|=\sqrt{2})$ parts:
\begin{equation} \label{SingTrip}
\bar{s}(K_4,K_3)U^S(K_1,K_2,K_3)s(K_2,K_1)+\bar{t}_{\mu}
(K_4,K_3)U^A(K_1,K_2,K_3)t_{\mu}(K_2,K_1), 
\end{equation}
where $s$ and $t_{\mu}$ are the variables of annihilation of the
singlet and triplet states
 \begin{equation} \label{Sing}
s(K_2,K_1)\equiv \frac{1}{\sqrt{2}}\sum _{\sigma}\sigma{\Psi}_{\sigma K_2}
{\Psi}_{-\sigma K_1},
\end{equation}
\begin{equation} \label{Trip}
t_{0}(K_2,K_1)\equiv \frac{1}{\sqrt{2}}\sum _{\sigma }{\Psi}_{\sigma K_2}
{\Psi}_{-\sigma K_1}\hspace{10mm} ; \hspace{10mm} t_{\pm 1}(K_2,K_1)\equiv 
{\Psi}_{\uparrow ,\downarrow K_2}
{\Psi}_{\uparrow ,\downarrow  K_1}.
\end{equation}
All coupling functions in the equations 
(\ref{CSoper},\ref{par-anipar},\ref{SingTrip}) possess the symmetry related to 
momentum--exchange and time--inversion.
If ${\cal F}(K_1,K_2,K_3,K_4)$ 
is a coupling function two exchange operators 
can be defined as
\begin{equation}\label{defX1}
X{\cal F}(K_1,K_2,K_3,K_4)\equiv {\cal F}(K_2,K_1,K_3,K_4)
\end{equation}
and 
\begin{equation}\label{defX2}
\bar{X}{\cal F}(K_1,K_2,K_3,K_4)\equiv {\cal F}(K_1,K_2,K_4,K_3).
\end{equation}
The time inversion operator ${\cal T}$ is
\begin{equation} \label{Timerev}
{\cal T}{\cal F}(K_1,K_2,K_3,K_4)\equiv {\cal
F}(K_3,K_4,K_1,K_2).
\end{equation}
The symmetries of the coupling function are the time inversion symmetry
\begin{equation} \label{Timerevsym}
{\cal T}{\cal F}={\cal F}
\end{equation}
and the  exchange symmetry
\begin{equation} \label{Xsym}
X\bar{X}{\cal F}={\cal F}\; .
\end{equation}
Both symmetries can be easily checked for the coupling functions in 
expression (\ref{par-anipar}). We will see that all other 
couplings can be derived from $U_{\perp}$ only  and have the same symmetry 
properties upon $X$ and $\cal{T}$ operations.
It is easy to see that $\bar{X}{\cal F}=X{\cal F}$ if 
${\cal T}{\cal F}={\cal F}$: exchanging particles 1 and 2 
or particles 3 and 4 are exchanged is equivalent.

We want now to find the 
relations between the six coupling functions in  equations 
(\ref{CSoper},\ref{par-anipar},\ref{SingTrip}).
Using the Pauli principle one gets
\begin{equation} \label{trafo1}
U_{\parallel}=U_c+U_{\sigma}\; ,
\end{equation}
 \begin{equation} \label{trafo2}
U_{\perp}=U_c-U_{\sigma}-2XU_{\sigma}.
\end{equation}
Let's suppose $U_c$ and $U_{\sigma}$ to be two independent functions.
We can write them in the form
\begin{equation} \label{ind1}
U_c=\frac{1}{4}(2-X)U_1+U_2\; ,
\end{equation}
\begin{equation} \label{ind2}
U_{\sigma}=-\frac{X}{4}U_1 \; .
\end{equation}
If we now {\it choose} $U_1=U_{\perp}$  it follows from (\ref{trafo2}) that 
$U_2=0$. It means that the most general interaction 
can be written  in terms of a single function $U_{\perp}$, without 
losing generality.
The function $U_{\parallel}$ is also contained in $U_{\perp}$. 
Namely, from
two equal-spin electrons one can build only a triplet state 
(antisymmetric under $X$) so that
\begin{equation} \label{Upar}
U_{\parallel}=U^A, 
\end{equation}
while
\begin{equation} \label{Uper}
 U_{\perp}=U^A+U^S\; ,
\end{equation}
containing the singlet and the triplet interactions. 
$U^A$ and $U^S$ can be seen as the antisymmetric and symmetric parts 
of the same function. This function is simply $ U_{\perp}$.

We see that all coupling functions are contained in $ U_{\perp}$ which we 
call simply $U$ or $U_l$ to make its scale dependence explicit.
One thus have
\begin{equation} \label{CSinter}
U_c=\frac{1}{4}(2-X)U \; , \hspace{10mm} U_{\sigma}=-\frac{X}{4}U\; ,
\end{equation}
\begin{equation} \label{AS}
U^A=U_{\parallel}=\frac{1}{2}(1-X)U\; ,\hspace{10mm}
U^S=\frac{1}{2}(1+X)U\; .
\end{equation}

The effective coupling function for the renormalization of the AF
correlation function (eq. \ref{VSDW}) is obtained from the spin coupling
\begin{equation} \label{sponcoup}
V_l^{AF}(\theta_1,\theta_2)=4U_{\sigma l}(\theta_1,\theta_2,\tilde{\theta_1})
\; ,
\end{equation}
where we take only the $\theta$--dependence of the coupling functions 
into account. The angle $\tilde{\theta}$ is related to the angle $\theta$
in such a way that the momentum difference between the particles 
${\bf k}(\theta)$
and ${\bf k}(\tilde{\theta})$ is the perfect nesting vector 
$(\pm \pi,\pm \pi)$.
The coupling function for the CDW at ${\bf q}=(\pi,\pi)$ would be
\begin{equation} \label{ccop}
V_l^{CDW}(\theta_1,\theta_2)=4U_{cl}(\theta_1,\theta_2,\tilde{\theta_1})
\; .
\end{equation}

%\end{multicols}

\end{document}